\documentclass[a4paper,12pt]{article}

\usepackage{soul}
\usepackage[usenames,dvipsnames]{color}
\usepackage{jheppub}
\usepackage{amsmath, amssymb, slashed, epsf, color, graphicx, latexsym}
\usepackage{epsfig}

\begin{document}
\preprint{TIFR/TH/15-31}
\title{A Charged Membrane Paradigm at Large D}
\author[a]{Sayantani Bhattacharyya,} 
\author[b] {Mangesh Mandlik,}
\author[b]{Shiraz Minwalla,}
\author[b]{and Somyadip Thakur}
\affiliation[a]{Indian Institute of Technology Kanpur, Kanpur, India-208016}
\affiliation[b]{Tata Institute of Fundamental Research, Mumbai, India-400005}
\emailAdd{sayanta@iitk.ac.in}
\emailAdd{mangesh@theory.tifr.res.in}
\emailAdd{minwalla@theory.tifr.res.in}
\emailAdd{somyadip@theory.tifr.res.in}

\abstract{
We study the effective dynamics of black hole horizons in Einstein-Maxwell 
theory in a large number of spacetime dimensions $D$. We demonstrate that 
horizon dynamics may be recast as a well posed initial value problem 
for the motion of a codimension one non gravitational membrane 
moving in flat space. The dynamical degrees of freedom of this 
membrane are its shape, charge density and a divergence free 
velocity field. We determine the equations that govern membrane dynamics 
at leading order in the large $D$ expansion. Our derivation of the membrane
equations assumes that the solution preserves an SO$(D-p-2)$ isometry 
with $p$ held fixed as $D$ is taken to infinity. However we are able 
to cast our final membrane equations into a completely geometric form 
that makes no reference to this symmetry algebra.}

\maketitle


\section{Introduction}\label{intro}

Emparan, Suzuki, Tanabe (EST) and collaborators have recently noted
\cite{Emparan:2013moa, Emparan:2013xia, Emparan:2013oza, Emparan:2014cia, 
Emparan:2014jca, Emparan:2014aba, Emparan:2015rva}
that the  classical dynamics of black holes simplifies at large 
$D$ ($D$ is the dimensionality of space time). Schwarzschild black holes 
in a large number of dimensions are characterized by two widely separated 
length scales. The first of these is the Schwarzschild radius $r_0$, while 
the second is the distance $\delta r$ away from Schwarzschild radius after 
which spacetime ceases to be warped by the black hole. In other words 
$\delta r$ is defined so that spacetime is effectively flat for 
$r > r_0+\delta r$. At large $D$ the membrane 
thickness,  $\delta r$, is easily estimated; it turns out that 
$\delta r \sim {r_0}/{D} \ll r_0$. Similar observations apply to 
static charged black holes at large $D$.

The separation of scales between the membrane thickness and the black hole
radius results in the simplification of black hole dynamics at large $D$. 
The first hint of this fact appeared in the results for the large $D$ 
spectrum of quasinormal modes of Schwarzschild black holes
obtained by EST and collaborators \cite{Emparan:2014cia, 
Emparan:2014aba, Emparan:2015rva}.  It turns out that most  
of the quasinormal modes  are heavy with frequencies 
$ \sim {1}/{\delta r}$. The remaining modes 
are anomalously light; their frequencies are of order ${1}/{r_0}$.
\footnote{More precisely, all but a finite number 
of quasinormal modes at every angular momentum are heavy. A finite number 
of modes at every angular momentum are light.}
As we will see below, the spectrum of quasinormal modes about 
Reissner-Nordstrom black holes is qualitatively similar.

The pattern of the quasinormal mode frequencies described above may 
be understood intuitively as follows. A quasinormal mode is a linearized solution of Einstein's 
equations about the black hole background, subject to the condition that it 
is ingoing at the horizon and outgoing in the asymptotically flat exterior 
region. As the second boundary condition is effectively imposed at the 
outer edge of the membrane region, the quasinormal problem is analogous to 
the analysis of the harmonics of the wave equation in a hollow, leaky 
spherical shell. The radius of this shell is $r_0$ and its 
thickness is $\delta r$. Clearly modes with nonzero `harmonic number' in the 
radial direction all have frequencies of order ${1}/{\delta r}$; 
these are EST's generic heavy modes. Modes of zero radial harmonic number, 
if present, have frequencies of order ${1}/{r_0}$; these are 
EST's anomalously light modes. 

The imaginary part of all heavy quasinormal mode frequencies are of order 
${1}/{\delta r}$; it follows that these modes all decay away after a 
time scale of order $\delta r$. On the other hand the light quasinormal modes 
have lifetimes of order $r_0$. Consider a violent dynamical process like a 
black hole collision. For a time of order $\delta r$ after the event, dynamics
is complicated and involves all quasinormal modes. For times 
$t \gg \delta r$, however, the heavy quasinormal modes have all decayed away
and the subsequent dynamics is  governed by a nonlinear interacting theory 
of only the light quasinormal modes, the principal focus of this paper.
\footnote{At time scales large compared to 
$r_0$ the light quasinormal modes also decay away and the black holes 
settle down into their equilibrium state. The approach to equilibrium is 
governed by the linearized theory of quasinormal modes.}

Light quasinormal modes may roughly be thought of as `Goldstone bosons' 
for the symmetries of flat space that are broken by the black hole.  
Non rotating black holes appear in family of solutions labeled by a set 
of parameters $\alpha^i$; the black hole location, radius, boost velocity and 
charge. By infinitesimally varying each of these parameters we 
obtain a set of time independent linearized solutions  
of the Einstein-Maxwell equations about any of these black holes. 
Now consider configurations that locally resemble these zero modes 
but with $\delta \alpha^i = \delta \alpha^i(\theta)$, i.e. with the 
infinitesimal parametric variations chosen to be functions of the black hole 
angular coordinates with spherical harmonic numbers of order unity. 
It follows that in any patch of size of order $\delta r$ (i.e. of 
angular extent of order $ \delta r /r_0$) the $\delta \alpha^i$ are 
approximately constant. In any such patch the fluctuation closely 
approximates a zero mode, and so is static on the time scale 
$\delta r$. However the variation of $\delta \alpha^i$ on 
length scales of order $r_0$ cause 
such configurations evolve over times of the same order. It follows that 
quasinormal modes built out of such configurations 
have frequencies of order ${1}/{r_0}$,  and may be identified with 
charged generalizations of the light modes of EST.

The identification of light quasinormal modes with `Goldstone bosons' 
immediately suggests the possibility of using the collective coordinate 
method to derive the {\it nonlinear} `chiral Lagrangian' of these light 
modes. \footnote{As the resulting system turns out to be dissipative, however, 
 it is easier to deal with the effective equations of motion than an 
effective action. }. On general grounds one expects that the 
effective nonlinear equations of motion for the light modes will admit 
a power series expansion in the ratio of the energy scales 
of the light and heavy modes, i.e. in a power series in 
$\delta r/{r_0} \sim {1}/{D}$. In other words the collective 
coordinate equations for light modes dynamics are a reformulation of black 
hole dynamics that is exact at large $D$.  
 
At leading nontrivial order in $1/D$, the equations that govern the collective 
coordinate dynamics of uncharged black holes were derived 
in the recent paper \cite{our} (see \cite{Emparan:2015hwa, Suzuki:2015iha, 
Emparan:2015gva, Suzuki:2015axa, Tanabe:2015hda} for 
closely related work) \footnote{The papers \cite{Emparan:2015hwa} and
\cite{Suzuki:2015iha} worked out the effective collective coordinate 
expansions for the special case of uncharged stationary configurations. When 
restricted to flat space and lowest order in $D$ the results of these 
papers are special cases of \cite{our} and 
the current paper.  The 
papers \cite{Emparan:2015gva, Suzuki:2015axa, Tanabe:2015hda} analyze 
dynamics at length and time scales of order $r_0/\sqrt{D}$ (this turns out 
to be the relevant length scale for the Gregory-Laflamme phenomenon at 
large $D$), as opposed to the current paper where we focus on length scales
of order unity.}. In this paper we 
build on the work of \cite{our} in two different ways. First we 
improve the construction of \cite{our} in several respects. We 
use collective coordinate variables with a direct physical significance 
and present our final equations and spacetimes in an explicitly 
`geometrical' form. Second - using the same improvements - 
we generalize the work of  \cite{our} to obtain the nonlinear collective 
coordinate dynamics of {\it charged} black holes in a large number of 
dimensions. 
 
In the rest of this introduction we will provide a more detailed description 
of the collective coordinate construction presented in this paper and 
present our main results. 
\\

{\bf A more detailed introduction and summary} 
\\

In the technical heart of this paper we follow \cite{our} to simply 
write down a class of leading order collective coordinate spacetimes 
(see \eqref{ansatz} below). 
We then carefully  verify that our spacetimes and gauge fields (written down by 
physically guided guesswork following \cite{our}) obey the Einstein-Maxwell 
equations of motion at leading order in ${1/D}$ \footnote{More precisely 
the equations of motion are obeyed everywhere outside the even horizons of 
these configurations. This is sufficient, as regions inside the event horizon 
are causally disconnected - and invisible - from those outside, and 
so may be ignored for the purposes of predicting observations outside 
the event horizon.} and so constitute 
a good starting point for the construction of true solutions to 
the Einstein-Maxwell equations in an expansion in $1/D$. Our 
collective coordinate spacetimes are built sewing together 
patches of Reissner-Nordstrom black holes with different radii, charges and boost velocities into 
a single smooth spacetime. These spacetimes 
are in one to one 
correspondence with the configurations of a non gravitational 
codimension one membrane propagating in flat $D$ dimensional space. 
The dynamical degrees of freedom of the membrane are
\begin{itemize}
\item{1.} The embedding of its timelike world volume in flat $D$ dimensional 
spacetime, i.e. the shape of the membrane. Through this paper we use the 
symbols $n_A$ and $K_{AB}$ to denote 
the normal and extrinsic curvature of the membrane surface in $D$ dimensional 
Minkowski space. We also use the symbol ${\cal K}=\eta^{AB}K_{AB}$ to denote 
the trace of the extrinsic curvature. 
\item{2.} A velocity vector field $u^A$ in the membrane world 
volume (so that $u\cdot n=0$) whose world volume divergence vanishes 
(i.e. $\nabla\cdot u=0$ where $\nabla$ is the covariant derivative on the 
membrane world volume). The velocity field is normalized in the usual manner 
$u\cdot u=-1$.  
\item{3.} A scalar charge density field $Q$ 
\footnote{More precisely the field $Q$ utilized in this paper is 
a variable proportional to the actual conserved charge density field 
on the membrane; see the upcoming paper \cite{bh} for details.}
that lives 
on the membrane (this field is absent in the neutral case). 
\end{itemize}

To reiterate, the starting point of the technical analysis presented 
in this paper is a class of `collective coordinate spacetimes' - that are
simply guessed. We have one such spacetime for every 
distinct membrane configuration. Our collective coordinate spacetimes  
turn out to solve the Einstein-Maxwell equations at leading order in $1/D$
everywhere outside their event horizons. 
 
The strategy adopted in the 
rest of this paper is to use these spacetimes as the first term in the 
perturbative construction of true solutions of the Einstein-Maxwell 
equations in a power series expansion in $1/D$. In this paper we 
 explicitly implement 
this expansion to first subleading order in $1/D$. In other words we 
correct the leading order collective coordinate spacetimes described 
above to ensure that they obey the Einstein-Maxwell equations not just at the 
leading order in $1/D$ but also at first subleading order in this expansion. 
We discover that it is possible to accomplish this task with 
only nonsingular corrections if and only if the membrane shape, charge 
density
and velocity fields obey the following local equations of motion 
\begin{equation}\label{Memeqsc}
\begin{split}
&\left(\frac{\nabla^2u}{\mathcal{K}}-(1-Q^2)\frac{\nabla\mathcal{K}}{\mathcal{K}}+u\cdot K - (1+Q^2)(u\cdot\nabla) u\right)\cdot{\cal P}=0,\\
&\frac{\nabla^2Q}{\mathcal{K}}-u\cdot\nabla Q - Q\left(\frac{u\cdot\nabla\mathcal{K}}{\mathcal{K}}-u\cdot K\cdot u\right)=0,\\
\text{where }&\nabla = \text{the covariant derivative 
on the membrane world volume,}\\
\text{and} ~~&{\cal P}_{AB} = \eta_{AB} - n_A n_B+ u_A u_B .
\end{split}
\end{equation}
\footnote{The expression in the first bracket in the first of 
\eqref{Memeqsc} is a vector in the membrane world volume and so is 
orthogonal to $n$. When acting on such a vector the projector 
${\cal P}_{AB}=g^{(WV)}_{AB}+u_A u_B$ where $g^{(WV)}_{AB}$ is the induced 
metric on the membrane world volume.}
\footnote{In the uncharged limit, the equation \eqref{Memeqsc} are easily demonstrated 
to reduce to the membrane equation of motion presented in \cite{our} once 
we account for the fact that the velocity field of this paper differs from 
the velocity field employed in \cite{our}
(see subsection \ref{relold} for relevant details.).}

Corresponding to every solution of the equations \eqref{Memeqsc} we are 
able to improve \eqref{ansatz}. The improvements are computed to ensure 
that the corrected configurations 
(see \eqref{metgform},\eqref{gaugegeo}, \eqref{metricgeo1}, 
 \eqref{metricgeo2}, \eqref{fdef})  solve the 
Einstein-Maxwell equations at leading {\it and first subleading} 
order in $1/D$. We expect the construction presented in this paper to 
constitute  the first couple of terms in a systematic expansion of 
solutions to the Einstein-Maxwell equations order by order 
in $1/D$. 

As we have explained above, membrane spacetimes are parameterized by the shape 
of the membrane (one function), the charge density field (one function) and 
a unit normalized divergence free velocity field on the membrane 
($D-3$ functions) and so by $D-1$ functions in total. 
The membrane equations \eqref{Memeqsc} are also $D-1$ in number 
(the first equation in \eqref{Memeqsc} is a vector projected orthogonal 
to $n$ and $u$ and so has $D-2$ components, while the second is a scalar 
and so has one component). It follows that we have as many equations as 
variables and so \eqref{Memeqsc} define an 
initial value problem for membrane motion. \eqref{Memeqsc} are simply the 
large $D$ collective coordinate equations of black hole motion.

Following \cite{our}, in this paper we have derived the membrane equations 
\eqref{Memeqsc} {\it under the assumption} that our spacetimes 
preserve an SO$(D-p-2)$ isometry subgroup
for $p$ held fixed as $D \to \infty$. \footnote{This requirement guarantees
that there are no unaccounted for factors of $D$ in, for instance, 
derivatives of the metric and gauge field. } Even though have made this 
assumption in our derivation, the final membrane equations 
\eqref{Memeqsc} (and the spacetimes dual to solutions of these membrane 
equations) make no explicit reference to the isometry group. Our final 
equations are entirely covariant; they treat the isometry directions and 
other directions democratically. We refer to equations with this 
property as geometrical.

Given the geometrical nature of our membrane equations and spacetimes, 
it is natural to wonder whether our equations apply more generally 
than their derivation. Could it be that \eqref{Memeqsc} captures the dynamics 
of black hole motions on time scales of order unity, even in the absence of 
a large isometry symmetry? While an appropriate version of such a 
conjecture might well be true, we would like to emphasize a subtlety. 
There are several pairs of independent geometrical expressions that 
reduce to each other at leading order in the large $D$ limit under 
the assumption of 
an SO$(D-p-2)$ isometry but differ from each other more generally 
\footnote{For example, the independent geometrical 
quantities 
$u\cdot  \nabla {\cal K}/D$ and $\nabla_\mu ( u\cdot  K)^\mu$ may be shown to agree with 
each other at leading order in $1/D$ for any membrane configuration that 
preserves an SO$(D-p-2)$ invariance.  On the other hand the same two 
expressions could differ at leading order when evaluated on configurations
that do not enjoy any symmetry. }.  For this reason it turns out that there
are different geometrical ways of presenting the equations of motion 
\eqref{Memeqsc}, all of which are identical at leading order in $1/D$ 
when evaluated on any membrane configuration that preserves an SO$(D-p-2)$ 
isometry but which differ on more general configurations. As the results of 
this paper are all obtained assuming an SO$(D-p-2)$ isometry, they cannot 
distinguish between these different geometrical presentations of the 
membrane equations. For example, the divergence of the first equation in 
\eqref{Memeqsc} turns out to coincide, at leading order in large $D$,
 with the equation 
\begin{equation}\label{sceq}
\begin{split}
&(1-Q^2)\left[\frac{{\nabla}^2\mathcal{K}}{\mathcal{K}^2}-\frac{u\cdot {\nabla}\mathcal{K}}{\mathcal{K}}\right] - (1+Q^2)\left(\frac{u\cdot {\nabla}\mathcal{K}}{\mathcal{K}}-u\cdot K\cdot u\right)=0,\\
\text{where }&\nabla = \text{the covariant derivative 
on the membrane world volume,}
\end{split}
\end{equation}
under the assumption of SO$(D-p-2)$ symmetry. It follows that the computations
presented in this paper cannot resolve the question of which of these is the 
`correct' leading order membrane equation in the absence of an isometry
.\footnote{Even staying within 
the class of isometric spacetimes, the iteration of the computations of this 
paper to one higher order could help to resolve this question. We 
hope to report on the results of a higher order computation in the not too  
distant future.}

The membrane equations \eqref{Memeqsc} are nonlinear and rather complicated. 
In an upcoming paper \cite{bh} we demonstrate that these equations admit 
simple classes of solutions in which the membrane velocity field $u^\mu$ is 
that of rigid rotations and the charge density field is proportional to 
$u^0$ (the time component of the velocity vector). The membrane shape 
is constrained to obey a single nonlinear partial differential equation.
Solutions obtained in this manner include the duals to charged rotating black 
hole solutions at large $D$. For the special case of uncharged black holes 
this nonlinear partial differential equation turns out to exactly match the 
constraint on the shape of stationary membranes 
derived in a different way in \cite{Emparan:2015hwa, Suzuki:2015iha}, 
establishing that the results of  \cite{Emparan:2015hwa, Suzuki:2015iha} 
(at leading order and in flat space) are a special case of the more general 
results of \cite{our} and the current paper. 

The simplest solution of the sort described in the previous paragraph 
is obtained upon switching off all angular velocities; the membrane 
solution is a  static spherical `soap bubble' with a uniform charge 
density. In section \ref{rns} below we have verified that the metric and gauge 
field dual to this solution agree perfectly with the exactly known static 
Reissner-Nordstrom black hole solution expanded to first subleading order 
in $1/D$ (we repeat this check for the much more nontrivial 
case of {\it rotating} black holes in 
the upcoming paper \cite{bh}). 

The membrane equations \eqref{Memeqsc} capture all of the complexities 
of black hole horizon dynamics at large $D$, at time scales of order unity 
\footnote{We believe this to be true at least for spacetimes that preserve 
an SO$(D-p-2)$ isometry for any $p$ that is held fixed as $D$ is taken to 
infinity.}. The detailed study of \eqref{Memeqsc} should teach us a great deal 
about black hole horizon dynamics. As a first 
small step in this program, in section \eqref{qnm}  we linearize 
the membrane equations  \eqref{Memeqsc} about the exact spherical solution 
dual to the Reissner-Nordstrom black hole, and determine the spectrum 
of small fluctuations about this background (see section \ref{rns}) for 
details. This spectrum of linearized fluctuations may be regarded as a 
prediction for the spectrum of light quasinormal modes 
about charged black holes at large $D$.

In the course of obtaining the quasinormal mode spectrum described in the 
previous paragraph, we reduce the manifestly geometrical but slightly 
abstract equations \eqref{Memeqsc} to explicit linear differential equations 
for two scalar fields and a divergence free vector field on 
$S^{D-2}$ times time (this reduction is valid for linearized fluctuations about
the spherical membrane surface). This explicit form of the equations helps us 
verify that the equations \eqref{Memeqsc} do indeed constitute a well posed
initial value problem for the membrane shape, charge density and velocity 
fields at least for these linearized configurations, as we had anticipated 
above on intuitive grounds. Our explicit results for the quasinormal modes 
also reveals that the membrane equations \eqref{Memeqsc} are highly 
dissipative. As an independent test of the equations \eqref{Memeqsc} it 
would be useful to verify our prediction for the large $D$ quasinormal spectrum 
by direct analysis of the Einstein-Maxwell equations about 
the Reissner-Nordstrom black hole background. While we make some 
remarks about this, we leave a detailed verification to future work.
\clearpage

\section{The collective coordinate ansatz}\label{detcolco}

\subsection{Boosted charged black holes in Kerr-Schild coordinates}\label{bcbhksc}

The Reissner-Nordstrom black hole in the `Kerr-Schild' coordinate system 
\footnote{See Appendix \ref{ks} for a lightning introduction to this 
coordinate system and its advantages.}
is given by 
\begin{equation}\begin{split}\label{bhgfks} 
ds^2&= -dt^2 + dr^2 + r^2 d \Omega_{D-2}^2 +
\left( (1+Q^2 c_D) \left( \frac{r_0}{r} \right)^{D-3} - 
c_D Q^2 \left( \frac{r_0}{r} \right)^{2(D-3)} \right) (dt +dr)^2,  \\
&=ds_{flat}^2 + \left( (1+Q^2 c_D) \left( \frac{r_0}{r} \right)^{D-3} - 
c_D Q^2 \left( \frac{r_0}{r} \right)^{2(D-3)} \right) (dt +dr)^2, \\
A&=\sqrt{2} Q \left( \frac{r_0}{r} \right)^{D-3} (dt+ dr).
\end{split}
\end{equation}

\eqref{bhgfks} describes a black hole at rest, i.e. a black hole moving 
with velocity $u=-dt$. The solution for a black hole 
moving at an arbitrary constant velocity $u$ may be obtained by boosting 
\eqref{bhgfks} and is given by 
\begin{equation}\begin{split}\label{bhgfkb} 
&g_{MN}= \eta_{MN}+
\left( (1+Q^2 c_D) \frac{1}{\rho^{D-3}} - 
c_D Q^2 \frac{1}{\rho ^{2(D-3)} }  \right)O_MO_N ,\\
&A_M=\frac{ \sqrt{2} Q O_M}{\rho^{D-3}} ,\\
&O=n-u, ~~~u = {\rm const}, ~~~u\cdot u=-1, ~~~
\rho= \frac{r}{r_0} \, ,\\ 
&r^2= P_{MN} x^M x^N, ~~~  
P_{MN}=\eta_{MN}+u_M u_N, ~~~{n}= r_0 d\rho, ~~~~{\rm note}~~u\cdot n=0\,.\\ 
\end{split}
\end{equation}

Note that the function $\rho$ in \eqref{bhgfkb} obeys the identity
\begin{equation}\label{bhradius}
\rho \nabla^2 \rho = (D-2) d\rho\cdot d\rho \, .
\end{equation}
Here and through most of this paper we view $\rho$ as a function that lives 
in flat $D$ dimensional space. In particular $\nabla$ in 
\eqref{bhradius} is the covariant derivative in flat space rather than 
in the metric \eqref{bhgfkb}.

Through this paper we will use the 
term membrane to refer to the surface $\rho=1$ viewed as a submanifold of 
flat Minkowski space.
Note also that $u^\mu$ may be thought of a vector field that lives 
on the membrane. It is obvious that 
\begin{equation}\label{vf}
\nabla\cdot u=0 ,
\end{equation}
where $\nabla$ is the covariant  derivative 
on the membrane.

\subsection{Collective coordinate spacetimes from boosted black holes} 
\label{ssansatz}

Consider the spacetime given by 
\begin{equation}\begin{split}\label{ansatz} 
g_{MN}&= \eta_{MN}+
\bigg[ \left(1+Q^2 \right) \frac{1}{\rho^{D-3}} - 
 \frac{Q^2}{\rho ^{2(D-3)} }  \bigg] O_M O_N  ,\\
A_M&=\frac{ \sqrt{2} Q O_M}{\rho^{D-3}} \,,\\
O&=n-u, ~~~u\cdot u =-1, ~~~{n} =\frac{d \rho}{\sqrt{d \rho. d \rho}}, ~~~u\cdot n=0, \\ 
\end{split}
\end{equation}
where $\rho$, $Q$ and $u$ are {\it arbitrary} smooth  
functions and vector fields in flat $D$ dimensional Minkowski spacetime 
subject only to the requirement that the function $\rho$ obeys 
\eqref{bhradius} on the membrane surface and that the velocity field 
restricted to the membrane obeys \eqref{vf}. 

The codimension one membrane worldvolume will play a special role in 
this paper. We assume  that the function $\rho$ is 
chosen to ensure that the membrane surface is a smooth timelike submanifold 
of flat Minkowski space. \footnote{We will 
see below that the same surface - $\rho=1$ - is a null when viewed as a 
submanifold of the metric \eqref{ansatz}.} The membrane 
separates regions of spacetime where with $\rho<1$ (inside the membrane) 
from regions with $\rho>1$ (outside the membrane). The function 
$\rho$ is chosen to ensure that the outside region is a connected 
spacetime and that includes all of spacelike infinity as well as ${\cal I}^+$ 
and ${\cal I}^-$. The membrane worldvolume itself is not necessarily 
connected.

The spacetimes \eqref{ansatz} have the following properties.  
\begin{itemize}
\item{1.} Upto corrections of order $1/D$, the static black holes 
\eqref{bhgfkb} are special cases of \eqref{ansatz} with the 
$\rho$, $Q$ and $u$ functions given as in \eqref{bhgfkb}. In these 
special cases $\rho=1$ is the black hole event horizon. 
\item{2.} It is easily verified  the membrane surface $\rho=1$ is a null 
submanifold of the metric \eqref{ansatz} for a general spacetime of this form.
At least when \eqref{ansatz} settles down to a stationary black hole at 
late times (as we will assume throughout this paper) this submanifold may be 
identified with the spacetime event horizon.  \footnote{The dissipative 
nature of the 
membrane equations of motion we derive below suggests that all solutions 
reduce to stationary solutions at late times.} 
\item{3.} Consider a point $x_0^\mu$ on the membrane $(\rho =1)$ of the 
spacetime \eqref{ansatz}. Let $u^\mu_0$, $Q_0$ and $\mathcal{K}_0$ denote the velocity, 
charge density field and trace of membrane extrinsic curvature at that point. 
Comparing with \eqref{bhgfkb}, we will see in subsection 
\ref{pertpatch} below  that a patch of size of order 
$\frac{1}{D}$ centered about $x_0^\mu$ is identical, at leading order in $D$, 
to the metric and gauge field of a patch centered about the membrane of 
a Reissner-Nordstrom black hole of radius $(D-2)/\mathcal{K}$, $Q$ parameter $Q_0$ and 
boost velocity $u^\mu_0$. 
\item{4.} It seems plausible from point (3) above that every patch centered 
about the membrane of the configuration \eqref{ansatz} obeys the 
Einstein-Maxwell equations at leading order in ${1}/{D}$. In subsection 
\ref{pertpatch} below we 
demonstrate that this is the case provided the spacetime \eqref{ansatz} 
enjoys an SO$(D-p-2)$ isometry for any $p$ that is held fixed as $D$ is 
taken to infinity.
\item{5.} The gauge field in \eqref{ansatz}  and the 
deviation of the metric from $ds_{flat}^2$ scales like $e^{-D(\rho-1)}$. 
It follows \eqref{ansatz} approaches flat space exponentially rapidly 
for $\rho-1 \gg {1}/{D}$. 
\item{6.} Combining (4) and (5) above it follows that
\eqref{ansatz} also obeys the Einstein-Maxwell equations at leading order 
in $1/D$ (or better) everywhere outside its event horizon. 
\item{7.} The equations of motion are not well solved when $1-\rho \gg 1$. 
However points that lie inside the event horizon of \eqref{ansatz} are 
causally disconnected from dynamics on and outside the membrane and will be 
ignored in the rest of this paper.
\end{itemize}

In summary, the metric \eqref{ansatz} is built 
by stitching together bits of the event horizon of Reissner-Nordstrom black
holes of varying radii, charge densities and boost velocities. The spacetime 
\eqref{ansatz} obeys the Einstein-Maxwell equations 
at leading order in large $D$ everywhere outside its horizon 
at least provided 
it preserves an SO$(D-p-2)$ isometry. 
It follows that metrics of the form \eqref{ansatz} are useful starting points 
for a perturbative construction of the solutions of the Einstein-Maxwell 
equation in an expansion in $\frac{1}{D}$.

\subsection{Subsidiary constraints on $\rho$, $u$ and $Q$} \label{sssubsid}

The spacetimes \eqref{ansatz} are parameterized by the functions 
$\rho$ and $Q$ and $u^\mu$. These functions are defined on all of $D$ 
dimensional Minkowski space. However we have already noted that 
\eqref{ansatz} rapidly tends to flat space when $\rho-1 \gg \frac{1}{D}$.
Consequently two spacetimes whose $\rho$ and $Q$ and $u^\mu$ functions 
agree on the surface $\rho=1$ but deviate at larger values of $\rho$ actually 
describe spacetimes that agree at leading order in $1/D$ 
on and outside their event horizons. \footnote{In and around 
subsection \ref{pertpatch} we show that for this statement to be 
true it is also necessary the gradients $\nabla \rho$ of the two $\rho$ 
functions coincide on the membrane $\rho=1$ at leading order in the 
large $D$ limit. However this is automatic, 
given the conditions we have imposed on our construction. Upto a position 
dependent normalization, $\nabla \rho$ is proportional to the normal 
vector of the surface $\rho=1$. It follows that the two $\nabla \rho$
functions agree with each other upto normalization at $\rho=1$. The 
condition that both $\rho$ functions obey \eqref{bhradius} at $\rho=1$ 
guarantees that the normalizations also agree at leading order 
in the large $D$ limit (see \eqref{effconstrho}). }

In this paper we use spacetimes of the form \eqref{ansatz} as 
the starting point for a perturbative expansion of true solutions of the 
Einstein-Maxwell system in a power series in $1/D$. Any two configurations 
of the form \eqref{ansatz} that 
differ from each other only at subleading orders in $1/D$ constitute 
equivalent starting points for perturbation theory. In order to 
restrict attention only to inequivalent configurations 
 it is convenient to invent a set of rules that 
determine the functions $\rho$, $u$ and $Q$ everywhere in spacetime,
in terms of the shape of the membrane and the values of the 
velocity and charge density fields {\it on the membrane}. 
We refer to these arbitrary rules  as subsidiary constraints on the 
functions $\rho$, $Q$ and $u$. 

There is a great deal of freedom in the choice of subsidiary constraints. 
Two different choices of these conditions lead to the same solution at 
any given order in perturbation theory. The differences between the starting 
points in perturbation theory are compensated for by the differences in 
the results of the perturbative expansion. 
 
While all choices of subsidiary constraints are on equal footing in principle, 
in practice some choices (those that most accurately approximate the true 
eventual solutions) lead to simpler results in perturbation theory than others.
After experimenting with a few options we have chosen, in this paper, 
to impose the following subsidiary constraints on $\rho$, $u$ and $Q$:
\begin{equation}\label{properties}\begin{split}
&\rho \nabla^2 \rho = (D-2) d\rho\cdot d\rho,\\
&u\cdot u=-1,~~n\cdot u=0,~~  {\cal P}^{MN}\left[\left(n\cdot \nabla\right) u_M+ \left(u\cdot \nabla\right) n_M\right]=0,\\
& n\cdot\nabla Q=0,\\
\text{where}~~&n= \frac{d \rho}{\sqrt{d \rho\cdot d \rho}},~~ ~{\cal P}^{MN}=\eta^{MN} - n^M n^N + u^M u^N.\\
\text{and }&\nabla = \text{the covariant derivative 
in the embedding flat space.}
\end{split}
\end{equation}

Let us pause to comment on our choice of subsidiary constraints. 
Recall that it is an important element of our construction that 
\eqref{properties} is obeyed {\it on the surface $\rho=1$} 
(see \eqref{bhradius}). This is 
a physical requirement, independent of arbitrary choices of subsidiary
conditions. Our first subsidiary condition \eqref{properties} simply asserts 
that \eqref{bhradius} continues hold everywhere; even away from the membrane. 
This condition is sufficient to determine the function everywhere in terms 
of the shape of the membrane (i.e. solutions to the equation $\rho-1=0$).

The third condition
in \eqref{properties} asserts that $Q$ is defined off the membrane surface 
by parallel transporting it along integral curves of the normal vector 
$n \propto d \rho$. The second condition \eqref{properties} determines
$u$ in terms of its value on the membrane by specifying its evolution under 
parallel transport under the same integral curves. 
\footnote{The subsidiary constraints adopted in this paper are chosen  
to permit simple comparison with exact uncharged rotating black hole 
solutions, see \cite{bh} for details. These conditions imposed   
in this differ from the rather elegant geometrical subsidiary constraints 
imposed in \cite{our}. }

\subsection{Fixing coordinate and gauge invariance} \label{sscogi}

In the next section we will describe the perturbative procedure we will employ 
to correct the spacetime \eqref{ansatz} in order to obtain 
a spacetime that solves the Einstein-Maxwell equations upto first subleading 
order in $1/D$. In order to find an unambiguous solution to this problem 
we  need to fix 
coordinate redefinition and Maxwell gauge ambiguities. In this subsection we 
describe our choice of coordinates and gauge. 

Let the spacetime metric in the solutions described by this paper take 
the form
\begin{equation}\label{metform}
g_{MN} = \eta_{MN}+h_{MN},
\end{equation}
where $h_{MN}$ is given, at leading order, by \eqref{ansatz}. We 
fix coordinate redefinition ambiguity by imposing the condition
\begin{equation}\label{impcond}
O^Mh_{MN}=0,
\end{equation}
where 
\begin{equation}\label{Odef}
O=n-u,
\end{equation}
and all indices in \eqref{impcond} are raised and lowered using the flat 
metric $\eta_{MN}$. Using the fact that $O\cdot O=0$, it is easily verified 
that the leading order metric \eqref{ansatz} does indeed obey
\eqref{impcond}.
 
In a similar manner we fix the Maxwell gauge ambiguity by imposing the 
condition 
\begin{equation}\label{maxgauge}
O^M A_M=0.
\end{equation}
Note that \eqref{maxgauge} is obeyed at leading order (see \eqref{ansatz}).

Note that our choice of gauge depends on $O$, and so on $n$ and $u$, which, 
in turn, depend on the membrane shape and velocity field in the particular
solution under study. Our choice of gauge is somewhat analogous to 
a background field gauge in the study of gauge theories, or, more closely, 
to the gauges adopted in the study of the fluid gravity correspondence 
(see e.g. \cite{Bhattacharyya:2008jc, Erdmenger:2008rm, Banerjee:2008th, Bhattacharyya:2008mz, 
Rangamani:2009xk, Hubeny:2011hd}). 

Note also that the coordinate choice adopted in this paper differs in 
detail from that of \cite{our}. As is clear from the discussion of this 
section, the gauge adopted here is completely geometrical. This is 
not true of the gauge adopted in \cite{our}, which singles out the 
isometry direction as special. 

\subsection{Perturbation theory} \label{secpt}

In the next section we will implement a perturbative procedure that can be 
used to correct \eqref{ansatz} at first subleading order in $1/D$. 
Roughly speaking  we search for a metric and gauge field of the form 
\begin{equation} \label{pertmg} \begin{split}
g_{MN}&= \eta_{MN} + h_{MN},\\
h_{MN}&= \sum_{n=0}^\infty \frac{h_{MN}^{(n)}}{D^n}\,,\\
A_M&=\sum_{n=0}^\infty \frac{A_M^{(n)}}{D^n}\,,\\
h_{MN}^{(0)}&= O_M O_N \left[(1+Q^2)\rho^{D-3} - Q^2\rho^{-2(D-3)}\right],\\
A_M^{(0)}&=\frac{\sqrt{2} Q}{\rho^{D-3}}\,,
\end{split}
\end{equation}
and attempt to find the correction fields $h_{MN}^{(1)}$ and $A_M^{(1)}$ that 
ensure that the Einstein-Maxwell equations are satisfied not just at leading 
order but also at first subleading order in $1/D$. In order to technically 
implement this idea, it turns out to be very helpful to assume 
our solutions preserve a large isometry group, as we describe in 
detail in the next section

\section{Perturbation theory assuming SO$(D-p-2)$ invariance}

\subsection{Careful definition of the large $D$ limit}\label{intlard}

In the computational part of this paper we follow \cite{our} to take the 
limit $D \to \infty$ while preserving an SO$(D-p-2)$ symmetry with $p$ held 
fixed. We take the large $D$ limit while maintaining a large isometry 
subgroup so that we can reliably estimate the scaling with $D$ of 
all terms in the equations we encounter.

The requirement that our solutions preserve an isometry group is less 
restrictive than it first appears for two reasons. First, several spacetimes
of physical interest (e.g. those that describe classes of black hole 
collisions) indeed preserve large isometry groups. Secondly, although 
the derivation of the membrane equations that we present below assumes 
an SO$(D-p-2)$ isometry, we will see that all our  final equations are 
entirely geometrical on the membrane world volume; the isometry directions 
are not special in any way. In particular our final equations are 
independent of $p$. 

While none of our final results will depend on $p$, all intermediate 
computations are performed within a framework that explicitly 
preserves SO$(D-p-2)$ invariance. In order to perform computations 
we assume that our metric and gauge field take the form
\begin{equation}\label{metgonmink} \begin{split}
ds^2&= g_{\mu\nu}(x^\mu) dx^\mu dx^\nu + e^{\phi(x^\mu)} d \Omega_d^2~~,\\
A&= A_\mu (x^\mu)dx^\mu,\\
d&=D-p-3, ~~~\mu=1 \ldots p+3,\\
\end{split}
\end{equation}
where $g_{\mu\nu}$, $\phi$ and $A_\mu$ are all arbitrary
functions of the coordinates $x^\mu$ but are independent of the angular 
coordinates on the $S^d$ in \eqref{metgonmink}. 
\footnote{In the special case of flat space
\begin{equation}\label{metonmink}
ds^2=  \eta_{\alpha\beta} dw^\alpha dw^\beta + dS^2 + S^2 d \Omega_d^2 =\eta_{\alpha\beta} dw^\alpha dw^\beta + dz_M dz^M,
\end{equation}
where $z^M$ are the $d+1$ Euclidean coordinates built out of angular
coordinates on $S^d$ and the radial coordinate $S$.}
Under this assumption the $D$ dimensional Einstein-Maxwell equations 
effectively reduce to a $p+3$ dimensional Einstein-Maxwell system 
coupled to the  effective scalar field $\phi$.

\subsection{The Einstein-Maxwell equations in the SO$(D-p-2)$ invariant 
sector}\label{emelard}

In this paper we study solutions of the Einstein-Maxwell equations governed
by the Lagrangian
\begin{equation}\label{lag}
{\cal S}= \frac{1}{16 \pi G_D} \int \sqrt{-\tilde{g}} ~d^Dx ~
\left( \tilde{R} - \frac{F_{MN} F^{MN}}{4} 
\right),
\end{equation}
where 
\begin{equation} \label{fadef}
\begin{split}
F_{MN}&= \partial_M A_N - \partial_N A_M,\\
\tilde{R} &= \text{Ricci scalar in full D dimensional spacetime,}\\
\tilde{g} &= \text{Determinant of the metric in full D dimensional spacetime.}
\end{split}
\end{equation} 
\footnote{In \eqref{fadef} the gauge field $A_\mu$ and the metric 
$g_{\mu\nu}$ are both taken to be dimensionless while Newton's constant 
$G_D$ has length dimension $D-2$.}
We wish to focus attention on metrics and gauge fields of the form 
\eqref{metgonmink}.
In this section we will work out the effective dynamical equations for 
such configurations.

Substituting \eqref{metgonmink} into \eqref{lag} we find the effective Lagrangian
\footnote{Due to the presence of SO$(D-p-2)$ symmetry all the quantities depend only on {$w^{\alpha},S$} coordinates, while all the vectors (in particular, $A$) have components only in {$dw^{\alpha},dS$} directions. Hence when we go to the $p+3$ dimensional space, the $M,N$ indices are replaced by $\mu,\nu$.}
\begin{equation}\label{lagphi} \begin{split}
{\cal S}&= \frac{\Omega_d}{16 \pi G_D} \int \sqrt{-g} ~d^{p+3}x ~
e^{\frac{d\phi}{2}}\left(R+d(d-1)e^{-\phi}+\frac{d(d-1)}{4}(\partial\phi)^2-\frac{F_{\mu\nu}F^{\mu\nu}}{4}\right),\\
(\partial\phi)^2 &= g^{\mu\nu}(\partial_{\mu}\phi)(\partial_{\nu}\phi).
\end{split}
\end{equation}
Varying this Lagrangian we obtain the equations of motion  
\begin{equation}\label{effeq}
\begin{split}
&(d-1)e^{-\phi}-\frac{d}{4}(\partial\phi)^2-\frac{1}{2}\nabla^2\phi + \frac{1}{4(d+p+1)}F_{\mu\nu}F^{\mu\nu}=0,\\
&R_{\mu\nu}- \frac{d}{4}(\partial_{\mu}\phi)(\partial_{\nu}\phi)-\frac{d}{2}\nabla_{\mu}\nabla_{\nu}\phi - \frac{1}{2}F_{\mu\rho}{F_{\nu}}^{\rho}+\frac{1}{4(d+p+1)}F_{\rho\sigma}F^{\rho\sigma}g_{\mu\nu}=0,\\
&\nabla_{\mu}F^{\mu\nu} + \frac{d}{2}(\partial_{\mu}\phi )F^{\mu\nu} = 0,\\
&\text{where} ~~d = D-p-3,\\
&\text{and } \nabla = \text{covariant derivative taken w.r.t. the metric $g_{\mu\nu}$.}
\end{split}
\end{equation}

\subsection{Setting up the perturbative computation}\label{suc}

\subsubsection{Convenient coordinates for flat space}\label{mgsod1}

The metric \eqref{ansatz} is completely determined once we specify 
the two scalar fields $\rho$ and $Q$ and the vector field $u^\mu$. 
These fields live in {\it flat space} and are constrained to obey 
the equation \eqref{properties}.

The following coordinates for flat space 
\begin{equation}\label{fsm}
ds_{flat}^2=  \eta_{\alpha\beta} dw^\alpha dw^\beta + dS^2 + S^2 d \Omega_d^2~,~~~~
i=\{0,1,\cdots,p+1\},~~d = D-p-3.
\end{equation}
are particularly useful for studying SO$(D-p-2)$ invariant 
configurations. In these coordinates the requirement of 
SO$(D-p-2)$ isometry implies that $\rho$, $Q$ and $u$ are functions of 
$({\{w^\alpha,S\}\equiv\{x^\mu\} })$ only. Moreover $u_{\theta_i}=0$ in every 
angular direction $\theta_i$ on the $S^{d}$. 

\subsubsection{The perturbative expansion of SO$(D-p-2)$ invariant 
solutions}

Metrics and gauge fields that preserve an SO$(D-p-2)$ isometry can be 
parameterized in the form 
\begin{equation}\label{decomp}
\begin{split}
ds^2&= g_{\mu\nu}(S,w^\alpha) dx^\mu dx^\nu +S^2 e^{\delta\phi(S,w^\alpha)} d \Omega_d^2~~,\\
A_M dX^M &= A_\mu (S,w^\alpha) dx^\mu.
\end{split}
\end{equation}
Note that 
$$ \phi= \phi^0+ \delta \phi, ~~~\phi^0= 2 \ln(S).$$
($\phi^0$ is simply value of $\phi$ in flat space).

As explained around \eqref{metform}, in this paper we will expand the metric 
and gauge field in a power series expansion in $1/D$. \footnote{The 
central advantage
of the assumption of SO$(D-p-2)$ isometry is that the variables of the 
perturbation expansion are independent of $D$.} 
The schematic expansion \eqref{metform} takes the precise form
\begin{equation}\label{metaexp}
g_{\mu\nu} = \sum_{k=0}^\infty \left(\frac{1}{D}\right)^k g_{\mu\nu}^{(k)},
~~A_{\mu} = \sum_{k=0}^\infty \left(\frac{1}{D}\right)^k A_{\mu}^{(k)},~~
\delta\phi=\sum_{k=1}^\infty \left(\frac{1}{D}\right)^k \delta\phi^{(k)} .
\end{equation}

From \eqref{ansatz} we read off the leading values of $g_{\mu\nu}$ and $A_\mu$ 
\begin{equation}\label{leadval}
\begin{split}
g^{(0)}_{\mu\nu} dx^\mu dx^\nu &= \eta_{\alpha\beta} dw^\alpha dw^\beta + dS^2 + \left[(1+Q^2)\rho^{-(D-3} - Q^2\rho^{-2(D-3)}\right] (O_\mu dx^\mu)^2,\\
A_\mu^{(0)} &= \sqrt{2}Q\rho^{-(D-3)} O_\mu.\\  
\end{split}
\end{equation}

\subsubsection{More detailed parameterization of the first order 
corrections to the metric and gauge field}

 After imposing the gauge conditions \eqref{impcond} and \eqref{maxgauge}, 
the metric correction $g_{\mu\nu}^{(1)}$ and gauge field 
correction $A^{(1)}_\mu$ can  can be parameterized in terms of 6 unknown scalar, 
three unknown vector and one unknown tensor functions \footnote{The terms 
scalar, vector and tensor refer to the transformation properties of the 
fields under those rotations in the tangent space that leave $n$, $u$ 
and $dS$ fixed. See below for more details. }as  
\begin{equation}\begin{split}\label{uk}
g^{(1)}_{\mu\nu}  & = S_{(VV)} O_\mu O_\nu + 2 S_{(Vz)} 
O_{(\mu} Z_{\nu) } + S_{(zz)} Z_\mu Z_\nu 
+  S_{(Tr)} P_{\mu\nu}  \\
& + 2 {V^{(V)}}_{(\mu }O_{\nu)}+  2 {V^{(z)}}_{(\mu }Z_{\nu)} + T_{\mu\nu}, \\
& A^{(1)}_\mu= S_{(AV)} O_\mu + S_{(Az)} Z_\mu + V^{(A)}_\mu, \\
\end{split}
\end{equation}
where
\begin{equation*}
\begin{split}
&O= n-u,~~
Z= \frac{dS}{S} -\left(\frac{n\cdot dS}{S}\right) n,\\
&P_{\mu\nu} = \text{projector perpendicular to $u$, $n$ and $Z$},~~
P^{\mu\nu} T_{\mu\nu}=0.
\end{split}
\end{equation*}
The vectors ($V^{(V)}_\mu,~V^{(Z)}_\mu,~V_\mu^{(A)}$) and tensor ($T_{\mu\nu}$) above are all projected orthogonal to 
$O$, $n$ and $Z$ (the tensor $T_{\mu\nu}$ is also assumed to be traceless).

Let us now consider the corrections of the `dilaton' function $\delta \phi$. 
We see from \eqref{chdef} and \eqref{eeqsc} that $\chi = D (d\phi)$ 
appears in the equations of motion. Were $\phi$ to have \
an ${\cal O}\left(\frac{1}{D}\right)$ fluctuation $\delta\phi^{(1)}$, 
this term would contribute to the equations of motion at leading order, 
invalidating the fact that the starting metric \eqref{ansatz} solves the 
Einstein-Maxwell equations at leading order. For the same reason $\phi$ at 
${\cal O}\left(\frac{1}{D}\right)^2$,  contributes to the Einstein-Maxwell 
equations at ${\cal O}\left(\frac{1}{D}\right)$. It follows that 
$\delta\phi^{(2)}$ is an unknown function that contributes to the first order 
perturbative equations at the same order as the 6 scalars that appear in 
\eqref{uk}, and so will have to be determined together with these six 
functions in the computation of the first corrections to \eqref{ansatz}.

\subsubsection{Auxiliary embedding space}\label{auxiliary}

The coordinate system \eqref{fsm} describes flat $R^D$ as the `fibration'
of an $S^d$ over a $p+3$ dimensional base space with metric
\begin{equation}\label{auxspace}
ds_{flat}^2=  \eta_{\alpha\beta} dw^\alpha dw^\beta + dS^2=\eta_{\mu\nu}dx^\mu dx^\nu,~~~~x^\mu =\{w^\alpha,S\}.
\end{equation}
The radius of the fibred $S^d$ is given by the coordinate $S$. 

Under the assumption of SO$(D-p-2)$ symmetry, the membrane world volume 
can be thought of as a codimension one ($p+2$ dimensional)  surface in the 
base space together with the $d$ dimensional spheres fibred over each 
of the base points on this surface. More generally all the ingredients - 
the functions $\rho$, $u^\mu$ and 
$Q$ - that go into the construction of the seed metric \eqref{ansatz} 
can all be regarded as functions and vector fields on the base space - which 
then determine SO$(d+1)$ invariant functions and vector fields on all of 
$R^D$ in the obvious manner. 
This is the viewpoint we will adopt while doing the computations described 
in this section. This viewpoint is convenient because the auxiliary space
\eqref{auxspace} makes no reference to $D$. Once we formulate our 
perturbation theory in terms of fields propagating on the auxiliary 
space \eqref{auxspace}, all factors of $D$ in the equations are completely 
manifest, allowing for a clean formulation of large $D$ perturbation theory. 

The end result of the first stage of our computation (e.g. the results 
presented in \eqref{pertcomp}) are all presented in terms of covariant
derivatives of the field $\phi$, $u$ and $Q$ viewed as scalar and 
one-form fields that live in the base or auxiliary space \eqref{auxspace}.

It is important to note general expressions built out 
of covariant derivatives of SO$(D-p-2)$ invariant 
fields in the auxiliary space \eqref{auxspace}
do {\it not} agree with the corresponding expressions built out of 
covariant derivatives of the same fields in the metric 
\eqref{fsm} of the embedding space \footnote{Roughly speaking the difference comes about in terms 
involving expressions like $\Gamma_{SM}^M$ with $M$ summed over. This 
expression receives contributions from $M$ ranging over the angular 
directions of $\Omega_d$ in the case of \eqref{fsm} but not in the case 
of \eqref{auxspace}. }. In Appendix \ref{translation} we have 
explored the dictionary between covariant expressions in the full flat 
$D$ dimensional space and the auxiliary space. Using these translation
formulae, we are then able to rewrite our final results for the 
first order corrected metric and gauge fields in terms of full spacetime 
covariant derivatives of $\rho$, $u$ and $Q$. Our final results, 
presented in the next section, are given in this language, and 
turn out to be geometrical, in a sense we describe in 
detail below. 

\subsubsection{Constraints and Subsidiary conditions recast in auxiliary 
space}

As we have explained in the previous section our construction 
\eqref{ansatz} works provided the functions $\rho$ and $u$ obey 
the conditions \eqref{bhradius} and \eqref{vf}. The $\nabla^2$ 
in \eqref{bhradius} is a Laplacian in the full flat 
space \eqref{fsm}, while the $\nabla$ operator in \eqref{vf} is 
the covariant derivative on the membrane, viewed as a submanifold 
of the full flat space \eqref{fsm}. In order to use these conditions
in our computations below, we need to rewrite them in terms of 
covariant derivatives on \eqref{auxspace} and on the membrane 
world volume viewed as a submanifold of \eqref{auxspace}. 
\footnote{All computations in the paper \cite{our} were performed in 
the auxiliary space \eqref{auxspace}. The final results of 
\cite{our} were presented in this auxiliary space, without being 
reconverted to the full space. Note also that in the auxiliary space, because of our choice of coordinates, all Christoffel symbols vanish  and the covariant derivatives are same as partial derivatives.}

Depending on context, we will use the symbol ${\tilde \nabla}$ to 
denote the covariant derivative either in the base space \eqref{auxspace}
or on the membrane viewed as a submanifold of \eqref{auxspace}. As we have 
explained in Appendix \ref{translation}, 
\begin{equation}\label{utr}
\nabla\cdot u = (D-p-2)  Z\cdot u + {\tilde \nabla}\cdot u,
\end{equation}
(in this equation $\nabla$ is the covariant derivative of the membrane viewed 
as a submanifold of $R^D$ while ${\tilde \nabla}$ is the 
covariant derivative on the membrane viewed as a submanifold of 
\eqref{auxspace}). Here  
\begin{equation}\label{zdef} 
Z= \frac{dS}{S} -\left(\frac{n\cdot dS}{S}\right) n.
\end{equation}
Using the fact that \eqref{vf} is assumed to hold for our ansatz metrics 
it follows from \eqref{zdef} that  
\begin{equation}\label{effuconst}
 Z\cdot u= -\frac{{\tilde \nabla}\cdot u}{D-p-2}.
\end{equation}

In a similar manner the fact that  \eqref{bhradius} is assumed to 
hold on the membrane of \eqref{ansatz} implies that
\begin{equation} \label{effconstrho}
\begin{split}
&(D-p-2) \frac{dS\cdot  {\tilde \nabla} \rho}{S}+ {\tilde \nabla^2} \rho
= (D-2)d\rho\cdot d\rho.\\
&\text{where } \tilde{\nabla} =\text{the covariant derivative on the 
space \eqref{auxmet}.}
\end{split}
\end{equation}

In an entirely analogous manner, the subsidiary condition 
\eqref{properties} can be recast in terms of covariant derivatives 
in the auxiliary space \eqref{auxspace}.
\begin{equation}\label{proprep}\begin{split}
&(D-p-2)\frac{\rho}{S}dS\cdot  {\tilde \nabla} \rho+ \rho{\tilde \nabla^2} \rho
= (D-2)d\rho\cdot d\rho,\\
&u_\mu u^\mu=-1,~~~n_\mu u^\mu=0,\\
& \left(\eta^{\mu\nu} +u^\mu u^\nu-n^\mu n^\nu\right)\left[\left(n^\alpha{\tilde \nabla}_\alpha\right) u_\mu+ \left(u^\alpha{\tilde \nabla}_\alpha\right) n_\mu\right]=0,\\
& n^\mu{\tilde \nabla}_\mu Q=0,\\
\text{where}~~&n_\mu= \frac{{\tilde \nabla}_\mu \rho}{\sqrt{({\tilde \nabla}_\nu\rho)({\tilde \nabla}^\nu\rho)}},\\
\text{and } &\tilde{\nabla} =\text{the covariant derivative on \eqref{auxspace}.}
\end{split}
\end{equation}

\subsection{Zooming in on patches} 

In this subsection we will identify a scaling limit of distance 
scales that admits an interesting large $D$ limit. For this purpose 
we turn back to the Einstein-Maxwell equations specialized to the case 
of SO$(D-p-2)$ invariant configurations and note that derivatives of the 
scalar field $\phi$ appear in \eqref{effeq} 
with additional factors of $D$ as compared to terms with an equal number 
of derivatives of $g_{\mu\nu}$ or $A_\mu$. This observation (see \cite{our}) 
suggests that we will obtain one class of nontrivial solutions to these 
equations if we assume that $g_{\mu\nu}$ and $A_\mu$ vary on length scale 
${1}/{D}$, i.e. the length scale of $\delta r$ (see the introduction) 
while $\phi$ varies at the length scale unity (at least upto corrections 
that are subleading in $1/D$). Under this assumption the solutions we study are 
characterized by two widely separated length scales, exactly like the 
black holes described in the introduction. \footnote{See \cite{our} for 
a more detailed discussion of the rational behind choosing this 
scaling limit.}

In order to describe the large $D$ limit of solutions characterized by 
two different length scales ($1/D$ and unity)  we adopt the following 
procedure. We view our manifold as a union of patches,
each of size ${1}/{D}$. Each patch is centered around a particular 
coordinate $x^\mu_0$. In each such patch we work with the scaled 
coordinates, metric, connections and gauge fields
\begin{equation}\label{scalgmg}
\begin{split}
x^{\mu} &= x_0^{\mu} + \frac{\alpha^{\mu}(q^a)}{D},\\
G_{ab} &= D^2\times (\partial_a \alpha^\mu) ~(\partial_b \alpha^\nu) g_{\mu \nu},\\
{\cal A}_a&=D \times( \partial_a \alpha^\mu) A_\mu,\\
\end{split}
\end{equation}
where $\alpha^\mu$ are any convenient ($D$ independent) functions of the 
coordinates $q^a$. Note that $G_{ab}$ differs from $g_{\mu \nu}$ transformed 
to $q^a$ coordinates by the scale factor $D^2$. In the same way the 
gauge field ${\cal A}_a$ differs from $A_\mu$ transformed to the coordinates 
$q^a$ by a scale factor $D$. The scale factors are chosen to scale up 
distances and holonomies on the patch to order unity. We also find it 
convenient to define the one-form field 
\begin{equation}\label{chdef}
\chi_a \equiv D~\partial_a\phi = \alpha_a^{\mu}\partial_{\mu}\phi .
\end{equation}

Note that $\chi_\mu$ is of order unity and constant (to leading order in 
$1/D$) in scaled patch coordinates (see \cite{our} for more discussion). 
The equations of motion may be rewritten in terms of scaled quantities as
\begin{equation}\label{eeqsc}
\begin{split}
{\cal E}_{\phi}\equiv~&\left(\frac{d}{D}\right)\nabla_a \chi^a+ \frac{\chi^2}{2} -\frac{2(d-1)}{d}e^{-\phi} - \left[\frac{D^2}{2 d(D-2)}\right]F_{cd}F^{cd}=0,\\
{\cal E}_{ab}\equiv~&R_{ab} -\left(\frac{d}{D}\right)\left(\frac{\nabla_a \chi_b + \nabla_b \chi_a}{2}\right) -\left(\frac{d}{4D^2}\right)\chi_a\chi_b -\frac{1}{2} F_{ac}{F_b}^c  +g_{ab}\left[\frac{F_{cd}F^{cd}}{4(D-2)}\right] =0,\\
{\cal E}_a\equiv~&\nabla_aF^{ab}+\frac{d}{2D}\chi_aF^{ab}=0,\\
&\text{where } \nabla= \text{the covariant derivative w.r.t. metric $g_{\mu\nu}$.}
\end{split}
\end{equation}
All quantities (curvatures, Christoffel symbols, field strengths) 
in \eqref{eeqsc} are constructed out of the scaled metric $G_{ab}$  and scaled gauge 
field ${\cal A}_a$.\\

The variables in these equations are all assumed to be of order 
unity. All factors of $D$ in these equations are explicit, and 
so the equations \eqref{eeqsc} are easily expanded in a power series 
in $1/D$. At leading order, in particular, the equations reduce to 
\begin{equation}\label{eeqscl}
\begin{split}
{\cal E}_{\phi}|_\text{leading}\equiv~&\nabla_a \chi^a + \frac{\chi^2}{2} -2e^{-\phi} - \left[\frac{F_{cd}F^{cd}}{2 }\right]=0,\\
{\cal E}_{ab}|_\text{leading}\equiv~&R_{ab} -\frac{\nabla_a \chi_b + \nabla_b \chi_a}{2}  -\frac{1}{2} F_{ac}{F_b}^c =0,\\
{\cal E}_a|_\text{leading}\equiv~&\nabla_aF^{ab}+\frac{1}{2}\chi_aF^{ab}=0.
\end{split}
\end{equation}

In this paper we 
search for solutions of these equations in each patch of the manifold. 
We require that solutions in neighbouring patches agree with each other 
where they overlap. We will find solutions of our equations order by 
order in an expansion in $\frac{1}{D}$.  

\subsection{Choice of `patch coordinates'}\label{patchoice}

In this paper we will follow \cite{our} to implement perturbation in 
$1/D$ in a patch of size $\sim{\cal O}\left(\frac{1}{D}\right)$ centered 
around an arbitrary point $x^\mu_0$ on the membrane ($\rho=1$ surface). 
We will then sew together the results from each patch to obtain 
a global correction to the metric and gauge field in \eqref{ansatz}.

In order to set up the computation in any given patch, 
we need an explicit choice of local coordinates in each patch, 
i.e. an explicit choice of the coordinates $\sim\{y^a\} $  as defined in 
equation \eqref{scalgmg}. 

Having imposed SO$(D-p-2)$ invariance we have three distinguished one-form 
fields in each patch. These one-forms are $n(x^\mu_0)$, $u(x^\mu_0)$ and 
$Z(x^\mu_0)$. Note that from \eqref{effuconst} it follows that 
$$Z\cdot n  =0,~~Z\cdot O=-Z\cdot u = {\cal O}\left(\frac{1}{D}\right),$$

where `$\cdot$' denotes contraction with respect to flat metric.\\
Let $Y^i$ denote a set of $p$ one-form fields chosen so that 
$$Y^i\cdot Z= Y^i\cdot n=Y^i\cdot O=0, ~~~Y^i\cdot Y^j=\delta^{ij}.$$
There is, of course, a great deal of ambiguity in the precise details of 
the $Y^i$ fields that will play no role in what follows.

Let $\{x_0^\mu \}= \{w^\alpha_0,S_0\}$ represent a point on the membrane in the metric 
\eqref{ansatz}. We wish to focus on the patch of size of order 
$\frac{1}{D}$ around $x_0^\mu$. We set up a local coordinate system 
for this patch as follows.
\begin{equation}\label{rsc} \begin{split}
R&=D (\rho -1),\\
V&=D( x^\mu - x_0^\mu)O_\mu(x_0), \\
\frac{z}{S_0}&=D(x^\mu-x_0^\mu) Z_\mu(x_0),\\
y^i&=D(x^\mu-x_0^\mu)Y^i_\mu(x_0).\\
\end{split}
\end{equation}

\subsection{The perturbative metric in a patch}\label{pertpatch}

In these coordinates and at leading order in the $\frac{1}{D}$ 
expansion, the rescaled metric and gauge field \eqref{scalgmg} take the form
\begin{equation}\label{bbm} \begin{split}
ds^2 &=  2 \left(\frac{S_0}{n_S^0}\right) dR~dV- \left[1 -(1 + Q_0^2) e^{-R} + Q_0^2 e^{-2R}\right] dV^2 \\
&+\left[ \frac{1}{1-(n_S^0)^2} \right] dz^2 + \sum_{i=1}^p dy^i dy^i+ {\cal O}\left(\frac{1}{d}\right),\\
&e^\phi=S_0^2,\\
A&= \sqrt{2}~ Q_0 e^{-R} dV + {\cal O}\left(\frac{1}{d}\right),
\end{split}
\end{equation}
where $ Q_0 = Q(x^\mu_0),~~n_S^0 = (n\cdot dS)|_{x^\mu = x^\mu_0}$\,.\\
\eqref{bbm} describes a configuration that is translationally invariant 
in the coordinates $V$ $z$ and $y^i$ (but not in $R)$. We refer to 
\eqref{bbm} as the black brane metric. Notice that black brane metrics
are parameterized by $S_0$, $n_S^0$ and the charge $Q=Q_0$. Recall  
$r_0={S_0}/{n_S^0}$ is the radius of the static black hole whose patch, 
when blown up about a membrane point with $S=S_0$, yields the black brane 
metric \eqref{bbm}. 

It is easily directly verified that the black brane
configuration \eqref{bbm} solves the leading large $D$ equations 
of motion  \eqref{eeqscl}. 

After appropriate scaling the metric and gauge field fluctuation at first order in $\left(\frac{1}{D}\right)$  (see\eqref{uk})  takes the following form in the `patch coordinates'

\begin{equation}\begin{split}\label{ukc}
G^{(1)}_{ab }  dq^a dq^b& = S_{(VV)} dV^2 + 2 \left[\frac{S_{(Vz)} }{S_0}\right]
dV dz+ \left[\frac{S_{(zz)}}{S_0^2}\right] dz^2 
+  S_{(Tr)} dy^i dy^i \\
& + 2 V^{(V)}_idy^i dV+  2\left[\frac{ V^{(z)}_{i }}{S_0}\right]dy^i dz + T_{ij}dy^i dy^j,\\
\\
& {\cal A}^{(1)}_a dq^a= S_{(AV)} dV +\left[\frac{ S_{(Az)} }{S_0}\right]dz + V^{(A)}_i dy^i. 
\end{split}
\end{equation}

\subsection{The structure of perturbative equations at first order}\label{peo1str}

Let us begin the process of determining the correction to our metric 
and gauge field in a patch (centered about an arbitrary point on the membrane).
Upon plugging first order corrected metric and gauge field 
into the Einstein-Maxwell equations, we find that each of these 
equations takes the schematic form 
\begin{equation}\label{sfem}
H v^{(1)}= s^{(1)}.
\end{equation}
The term  $v^{(1)}$ in \eqref{sfem} is a schematic for the collection of 
unknown functions in \eqref{uk}. The `source' terms $s^{(1)}$ have their 
origin in the fact that a blown up patch of \eqref{ansatz} fails to solve 
the Einstein-Maxwell equations at first subleading order in $1/D$. This 
failure has its roots in the following facts:  
\begin{itemize} 
\item{1} A patch of \eqref{ansatz} differs from the black brane 
metric at first subleading order in $1/D$. This difference is visible 
upon Taylor expanding the fields $n$, $u$ and $Q$ to first order about 
the special point $x_0^\mu$ and results in source terms proportional to 
the first derivative of $n$ $u$ and $Q$.  
\item{2.} The black brane itself fails to solve the Einstein-
Maxwell equations at first subleading order in $1/D$. 
This shows up in the fact that the equations \eqref{eeqsc} themselves 
have corrections in the 
${1}/{D}$ expansion. This gives rise to derivative free source terms. 
\end{itemize}

Note that all source  terms are entirely determined by the 
data (membrane shape, velocity field, charge field) that go into defining 
the ansatz metric and gauge field. \eqref{ansatz}

All source terms are fast varying functions of the coordinate $R$ but slow 
varying functions of all other coordinates. This implies that 
$$v^{(1)}=v^{(1)}(R, \frac{V}{D}, \frac{z}{D}, \frac{y^i}{D}),$$ 
where $R$ and the other scaled coordinates are are defined 
in \eqref{rsc}. As $v^{(1)}$ is already a fluctuation variable at order $1/D$, 
derivatives of $v^{(1)}$ in all directions other than $R$ contribute to 
the Einstein-Maxwell equations only at order $1/D^2$. It follows that 
the homogeneous operator $H$ is a differential operator 
only in the variable $R$. In other words the equations \eqref{sfem}
are linear ordinary differential equations. 

Even though the RHS of \eqref{sfem} has its origin partly in the Taylor 
expansion of \eqref{ansatz} about the special point $x_o^\mu$, 
the source functions $s^{(1)}$ in the patch about $x_0^\mu$ do 
not explicitly depend on the expansion coordinates $V, z, y^i$. 
The reason for this is simple. The locality of the Einstein-Maxwell 
equations ensures that $s^{(1)}$ is a $\frac{1}{D}$ times a local functions of 
the fields $\rho^{p+d}$, $n^\mu$, $u^\mu$, $Q$ and their derivatives. 
Dependence on the coordinates 
$V$, $z$ and $y^i$ dependence could only arise from Taylor expanding the fields
$n^\mu$, $u^\mu$ and $Q$ about the point $x_0^\mu$. The terms proportional to 
$V, z,y^i$ in this Taylor expansion are all manifestly of order $1/D^2$ 
or smaller. \footnote{On the other hand source functions have nontrivial 
dependence on $R$ at leading order in $1/D$; this is a consequence of 
the fact that $\rho^{p+d}$ evaluates to $e^R$ at leading order in the large 
$d$ expansion, and so powers and derivatives of this function naturally 
appear in sources.}

Let us also reiterate that source $s^{(1)}$ contains at most one derivative 
of $n^\mu$, $u^\mu$ and $Q$. This follows immediately from the observation 
that $\rho$, $u$ and $Q$ are 
functions of $\frac{V}{D}$, $\frac{z}{D}$ and $\frac{y^i}{D}$ in the patch, 
and  every derivative of these functions is weighted by a factor of 
$\frac{1}{D}$. 

Let us summarize. \eqref{sfem} is a collection of 
an infinite number of {\it linear ordinary differential equations} in the 
variable $R$; one such equation at each point on the membrane world volume. 
At each membrane point the source functions are explicit function of $R$, 
with coefficients that depend on the values and (at most) one  
derivatives of the $\rho$, $u$ and $Q$ fields at that point. In to find 
$G^{(1)}_{ab}$, ${\cal A}^{(1)}_a$ and $\delta\phi^{(2)}$we need to solve these 
linear differential equations at each membrane point and then sew these 
solutions together into a global correction to \eqref{ansatz}.  
At the technical level, the procedure for perturbation theory is 
strongly reminiscent of the procedure adopted in studies of the fluid 
gravity correspondence, see e.g. \cite{Bhattacharyya:2008jc, 
Bhattacharyya:2008mz, Rangamani:2009xk, Hubeny:2011hd}

\subsection{Equations in 
the three symmetry channels}\label{symchan}

As we have explained above, the variables in $G^{(1)}_{ab}$, ${\cal A}^{(1)}_a$ and $\delta\phi{(1)}$ consist of 
$7$ scalar functions, 3 vector functions and one tensor function (where 
`scalar', `vector' and `tensor' refer to the transformation property of the 
modes under SO$(p)$ rotations in part of $x^\mu$ tangent space that is 
orthogonal to $X$, $n$ and $u$). The black brane background \eqref{bbm}), and 
so the operator $H$, preserves SO$(p)$ symmetry. It follows that the 
equations \eqref{sfem} do not mix the scalar vector and tensor modes; 
the equations in these three sectors decouple from each other. \\

{\bf Tensor Sector:}\\

In the tensor sector the differential equations \eqref{sfem} reduce to a
single ordinary second order differential equation for a single unknown, $T_{ij}(R)$; this equation is easily solved for an arbitrary source function. 
We present our explicit results below. \\

{\bf Vector Sector:}\\

In the vector sector we have four coupled equations for three unknown 
functions. The four equations in question are
\begin{equation}\label{fveq} \begin{split}
 &{\cal E}_{Ri} =0,
 ~~~ {\cal E}_{Vi} =0 ,\\
&  {\cal E}_{zi}=0 ,
 ~~~ {\cal E}_i =0 ,\\
\end{split}
\end{equation}
(see \eqref{eeqsc} for definitions of the equations). The directions $i$ 
are the $Y^i$ directions. They are assumed to be orthogonal to $O$, $u$ and 
$dS$. 

At first order it turns out that the following linear combination of equations vanishes identically.
\begin{equation}\label{lce}
\begin{split}
&\partial_R \bigg[\left(\frac{S_0}{n_S^0}\right){\cal E}_{Vi}  +f_0(R) {\cal E}_{Ri} \bigg] +\bigg[\left(\frac{S_0}{n_S^0}\right){\cal E}_{Vi}  +f_0(R) {\cal E}_{Ri} \bigg] +\left[\frac{1-(n_S^0)^2}{S_0} \right]  {\cal E}_{zi}=0,\\
&\text{where}~~ f_0(R) = 1- (1 + Q_0^2) e^{-R} + Q_0^2 e^{-2R}.
\end{split}
\end{equation}
We thus have only three independent vector equations for our three 
vector unknowns. It turns out that the the remaining 
three equations are easily solved for arbitrary source terms that 
obey \eqref{lce}, and in particular for the source terms that 
actually appear in the first order computation (see below for more details).\\

{\bf Scalar Sector:}\\

In the scalar sector we have 11 equations for 7 variables. The 11
 equations 
are
\begin{equation}\label{fseq} \begin{split}
& {\cal E}_{RR} =0 ,
 ~~~ {\cal E}_{RV} =0,
 ~~~ {\cal E}_{Rz} =0 ,\\
& {\cal E}_{VV}=0 ,
~~~  {\cal E}_{Vz}=0 ,
~~~  {\cal E}_{zz}=0 ,\\
& {\cal E}_{R}=0 ,
 ~~~  {\cal E}_{V}=0 ,
~~~{\cal E}_{z} =0 ,\\
& \sum_{i=1}^p{\cal E}_{ii} =0 ,
~~~ {\cal E}_{\phi}=0 ,\\
\end{split}
\end{equation}
(see \eqref{eeqsc} for the definition of these equations).
At first order it turns out that the following  four linear combination of 
equations automatically vanish.
\begin{equation}\label{lces} \begin{split}
\text{Combination-1:}~~&\partial_R{\cal E}_R +{\cal E}_R+\frac{{\cal E}_z}{S_0}=0,\\
\text{Combination-2:}~~&\partial_R\bigg[{\cal E}_{VV} +\left( \frac{n_S^0}{S_0}\right)f_0(R) {\cal E}_{RV}\bigg]+\bigg[{\cal E}_{VV} +\left( \frac{n_S^0}{S_0}\right)f_0(R) {\cal E}_{RV}\bigg] \\
&+\left(n_S^0 -\frac{1}{n_S^0}\right)\bigg[{\cal E}_{Vz}  + \frac{Q_0S_0e^{-R}}{\sqrt{2}\left[1 -( n_S^0)^2\right]}{\cal E}_R\bigg]=0,\\
\text{Combination-3:}~~&\partial_R\bigg[\left(\frac{n_S^0}{S_0} \right)f_0(R) {\cal E}_{Rz} + {\cal E}_{Vz}\bigg]+\bigg[\left(\frac{n_S^0}{S_0} \right)f_0(R) {\cal E}_{Rz} + {\cal E}_{Vz}\bigg]\\
&- \left(n_S^0 -\frac{1}{n_S^0}\right)\left(\frac{n_S^0}{S_0} \right){\cal E}_{zz} =0,\\
\text{Combination-4:}~~&\partial_R\bigg[{\cal E}_{\phi} + 2\left(\frac{n_S^0}{S_0}\right)^2 f_0(R)~ {\cal E}_{RR} - 2[1-(n_S^0)^2]~{\cal E}_{zz} -{\cal E}_{ii}\bigg] \\
 &+2\left(\frac{n_S^0}{S_0}\right)^2 \left[\partial_R f_0(R) + 2 f_0(R)\right]~{\cal E}_{RR}
 +4\left( \frac{ n_S^0}{S_0}\right){\cal E}_{RV} \\
 &+ 4\left[\frac{1-(n_S^0)^2}{S_0}\right] {\cal E}_{Rz}-2\sqrt{2} Q_0e^{-R}{\cal E}_{V}=0.
\end{split}
\end{equation}
We thus have exactly seven independent equations to solve for the seven 
unknowns in the scalar sector.  It turns out that the remaining seven 
equations are easily solved for arbitrary sources that obey \eqref{lces}, 
and in particular for the source terms that actually appear in the first 
order computation (see below for more details).

\subsection{Basis for Source Functions}\label{bsf}

Let us now turn to a description of the sources that appear on the RHS of 
\eqref{sfem}. In the scalar sector there are two kinds of sources. The 
first kind of source has its origin in the fact that the black brane 
metric \eqref{bbm} solves the Einstein-Maxwell equations only at large 
$D$ and not at first subleading order in $\frac{1}{D}$. This fact gives 
rise to sources (RHS of \eqref{sfem}) that are simply functions of $R$. 
We also have sources from the first term in the Taylor expansion of the 
functions $n_\mu$, $u^\mu$ and $Q$ expanded about $x_0^\mu$. Let 
${\mathfrak s}^{(a)}$ ($a = 1 \ldots N_S$) denote the set of scalar first derivatives 
of the functions $n$, $u$ and $Q$ . Let ${\mathfrak s}_0 =1$ (this allows us 
to deal with the first kind of source mentioned above). 
On general grounds, the source ${\cal S}_{m}$ terms in the $m$th scalar equations $E^S_{m}$ 
takes the form 
\begin{equation}\label{stesm}
{\cal S}_{m}= \sum_{a=0}^{N_S} {\cal S}_{m}^a(R)~ {\mathfrak s}^{(a)}.
\end{equation}

In a similar manner we let ${\mathfrak v}^{(a)}$ ($a = 1 \ldots N_V$) 
denote the set of scalar first derivatives of the functions $n$, $u$ and $Q$. 
The source terms ${\cal V}_{mi}$ in the $m$th vector equation $E^V_{mi}$ take the form 
\begin{equation}\label{vtesm}
{\cal V}_{mi}= \sum_{a=1}^{N_V} {\cal V}_{m}^a(R) {\mathfrak v}^{(a)}_i.
\end{equation}
Finally if ${\mathfrak t}^{(a)}_{ij}$ ($a = 1 \ldots N_T$) 
denote the set of tensor first derivatives of the functions $n$, $u$ and $Q$, 
then the source terms ${\cal T}_{ij}$ in the unique tensor equation must take the 
form 
\begin{equation}\label{ttesm}
{\cal T}_{ij}= \sum_{a=1}^{N_T} {\cal T}^a(R) {\mathfrak t}^{(a)}_{ij}.
\end{equation}

It turns out at first order $(N_S=6,~~N_V= 5,~~N_T=2)$. In table (\ref{table:1storder}) we have listed and explicit basis for independent scalar vector and 
tensor data at first order.  Here $P_{\mu\nu}$ is the projector perpendicular to $u_\mu$, $n_\mu$ and $Z_\mu$.
\begin{table}[ht]
\vspace{0.5cm}
\centering 
\begin{tabular}{|c| c|  c|} 
\hline 
Scalars &Vectors & Tensors \\
(6) & (5) & (2)\\ [1ex] 
\hline
\hline
${\mathfrak s}^{(1)} = u^\mu u^\nu K_{\mu\nu}$ & ${\mathfrak v}_\mu ^{(1)} = u^\nu P_\mu^{\alpha}K_{\nu\alpha}$ & ${\mathfrak t}_{\mu\nu}^{(1)} = P_\mu^\alpha P_\nu^\beta\left[\frac{K_{\alpha\beta}}{2} -\left(\frac{{\mathfrak s}_3}{p}\right)\eta_{\alpha\beta} \right]$\\
${\mathfrak s}^{(2)}= u^\mu Z^\nu K_{\mu \nu}$& ${\mathfrak v}_\mu ^{(2)} = u^\nu P_\mu^{\alpha}\partial_\nu u_\alpha$  &${\mathfrak t}_{\mu\nu}^{(2)} = P_\mu^\alpha P_\nu^\beta\left[\frac{\partial_\alpha u_\beta +\partial_\beta u_\alpha}{2} -\left(\frac{{\mathfrak s}_4}{p}\right)\eta_{\alpha\beta} \right]$  \\
${\mathfrak s}^{(3)} = P^{\mu\nu}  K_{\mu \nu}$& ${\mathfrak v}_\mu ^{(3)} =  P_\mu^{\alpha}(Z\cdot\partial) u_\alpha$ &\\
${\mathfrak s}^{(4)} = P^{\mu\nu}  \partial_\mu u_\nu$&${\mathfrak v}_\mu ^{(4)} =  P_\mu^{\alpha}\partial_\alpha Q$&\\
${\mathfrak s}^{(5)} = u^\mu \partial_\mu Q$& ${\mathfrak v}_\mu ^{(5)} =  Z^\nu P_\mu^{\alpha}K_{\nu\alpha}$ &\\
${\mathfrak s}^{(6)} = Z^\mu Z^\nu K_{\mu\nu}$&  &\\
\hline
\end{tabular}\vspace{.5cm}
\caption{Data at 1st order in $\frac{1}{D}$ expansion}
\label{table:1storder} 
\end{table}
\noindent

\subsection{Equations of motion from regularity at the horizon}\label{eomregh}

We are interested in solutions to the equations of perturbation theory 
that are everywhere regular (away from the black hole singularity that 
will turn out to be shielded by an event horizon). Even though all our 
source functions are regular, this condition is not automatic at 
$R=0$ (i.e. $\rho=1$). This perhaps surprising fact 
plays a key role in this paper. This subsection is devoted to a more 
detailed exposition of this fact. 

Let $E^{MN}$ denote the Einstein equation obtained by 
varying the Einstein-Maxwell Lagrangian w.r.t $g_{MN}$,  and let 
$M_N$ denote the Maxwell equation obtained by varying 
the Einstein-Maxwell Lagrangian w.r.t $A_M$. As we have explained above, 
the perturbative procedure of this paper is geared to determining the 
$\rho$ dependence of unknown metric and gauge field components. For 
our purposes it is thus natural to view the $\rho$ direction as a 
Euclidean `time' direction in which we wish to understand `dynamics'. 
From this point of view the equations 
\begin{equation}\label{consteqs}
C_{Ein}^M= E^{MN}(d\rho)_M= E^{M \rho} , ~~~C_{Max}= M^N(d\rho)_N = M^\rho,
\end{equation}
are, respectively, the Einstein and Maxwell `constraint' equations.

The dot product of the Einstein scalar equation $C_{Ein}^M$  with $n$ and $u$ 
(or $n$ and $O$) appears to play no role in the discussions of this subsection. 
For that reason in the rest of this section we will deal with 
$C_{Einp}^M$, the constraint Einstein equations that are projected 
orthogonal to $n$ and $O$. From the `geometrical' viewpoint (see below 
for much more discussion) $C_{Einp}^M$ is a vector equation while $C_{Max}$
is a scalar equation. However perturbative procedure described so far 
is not geometrical: it treats the isometry directions as special. 
From our current point of view $C_{Einp}^M$ may be decomposed into a single 
SO$(p)$ scalar $C_{Einp}\cdot Z$ and an SO$(p)$ vector ($C_{Einp}^M$ projected
orthogonal to $Z$). 

In the scalar sector it is easily verified that 
\begin{equation}\label{expfis1}
\begin{split}
\left(C_{Einp}\cdot Z\right) \propto &~\left[\left(\frac{S_0}{n^{0}_S}\right){\cal E}_{Vz} + f_0(R) {\cal E}_{Rz}\right],\\
\propto &~f_0(R)^2\frac{d}{dR}\left[\frac{S_{(Vz)}(R)}{f_0(R)}\right] + \Sigma_{(Vz)}(R)
=0,\\
\end{split}
\end{equation}

\begin{equation}\label{expfis2}
\begin{split}
C_{Max} \propto~ &{\cal E}_R\\
\propto~& f_0(R)\left( \frac{d}{dR} S_{(Az)}(R) \right)+\sqrt{2} Q_0e^{-R}S_{(Vz)}(R)+ \Sigma_{(Az)}(R)=0.\\
\end{split}
\end{equation}

Here $\Sigma_{(Vz)}(R)$ the full source term for the combination of equations 
$\left[\left(\frac{S_0}{n^{0}_S}\right){\cal E}_{Vz} + f_0(R) {\cal E}_{Rz}\right]$
while $\Sigma_{(Az)}(R)$ is the source term in ${\cal E}_R$. \footnote{
Clearly, each of 
$\Sigma_{(VZ)}(R)$ and $\Sigma_{(Az)}(R)$ are linear 
combinations of the previously defined quantities ${\cal S}^a_i(R)$.}

An inspection of \eqref{expfis1} reveals that this equation admits nonsingular 
solutions at $R=0$ if and only if the linear term,  in the Taylor expansion 
of ${\Sigma}_{(Vz)}(R)$ about $R=0$,  vanishes. Provided this condition is met 
the solution to \eqref{expfis1} is nonsingular. Once this condition 
is met it follows from  \eqref{expfis1} that 
\begin{equation}\label{svr}
S_{(Vz)}(R=0) =  \frac{\Sigma_{(Vz)}(R=0)}{f'(R=0)}\,.
\end{equation}

Turning to the equation \eqref{expfis2}, it is easily seen that the 
solution to this equation is nonsingular if and only if 
$\left[\sqrt{2} Q_0e^{-R}S_{(Vz)}(R)+ \Sigma_{(Az)}(R)\right]$
vanishes at $R=0$. Using \eqref{svr}, this condition is equivalent 
to the requirement that $\left[\sqrt{2} Q_0 \frac{\Sigma_{(Vz)}(R=0)}{f'(R=0)}+ \Sigma_{(Az)}(R=0)\right]$
vanish. Plugging in the explicit expressions for the source 
functions  $\Sigma_{(Az)}(R)$ and $\Sigma_{(Az)}(R)$
we find that we have nonsingular solutions if and only if 

\begin{equation}\label{relationss} 
\begin{split}
&(X-u)\cdot K\cdot (X-u) -\left[\frac{2Q^2}{(1-Q^2)}\right]\bigg[
(X-u)\cdot K \cdot u \bigg]=\left(\frac{1-n_S^2}{S~ n_S}\right),\\ 
&(X-u)\cdot \partial Q= Q\bigg[ (X-u)\cdot K\cdot u \bigg],\\
\text{where}~&X = \frac{dS}{n_S} -n = \left(\frac{n_S}{S} \right) Z.\\
\end{split}
\end{equation}

In the vector sector, the projection of $C_{Einp}$ may be shown to be 
proportional to 
\begin{equation}\label{expfiv}
\begin{split}
&\left[\left(\frac{S^0}{n_S^0}\right) {\cal E}_{Vi} +f_0(R)  {\cal E}_{Ri}\right]
\propto
f_0(R) \frac{d}{dR} \left[V^{(z)}_i(R)\right] +{\cal V}^{(Z)}_i(R)=0.
\end{split}
\end{equation}
Here ${\cal V}^{(Z)}_i(R)$ is the combination of source terms in the first 
line of \eqref{expfiv} - and so an appropriate linear combination of 
${\cal V}^a_{mi}(R)$
This equation has regular solutions if ${\cal V}^{(Z)}_i(R)$ vanishes at $R=0$
i.e. if 
\begin{equation}\label{relationsv}
 \begin{split}
&P_j^i \left[(X-u)\cdot \partial(u-n)_i + Q^2 \left( X\cdot \partial n_i - u\cdot \partial u_i \right) \right]=0,\\
\text{where}~&X = \frac{dS}{n_S} -n = \left(\frac{n_S}{S} \right) Z.\\
\end{split}
\end{equation}

It may be verified that \eqref{relationss} and \eqref{relationsv} exhaust 
the constraints of regularity; once these equations hold the 
solution for the first order correction to the black brane metric and 
gauge field can always be chosen (by choosing appropriate integration 
constants in the solutions of the differential equation) to be regular 
at $R=0$ (and everywhere else within the patch). 

In summary, the perturbative procedure described in this subsection yields 
regular solutions if and only if the equations of motion
\eqref{relationss} and \eqref{relationsv} are obeyed.

\subsection{Equivalence to the equations of \cite{our} in the uncharged limit} \label{relold}
Note that  the same null one-form $O_\mu$ has been parametrized in a different way in \cite{our}. 
\begin{equation} \label{oe}
O = A (dS - u^{there})= n-u^{here},
\end{equation}
where $u^{there}$ is the velocity field used  in \cite{our}  and in this subsection, $u^{here}$ will denote the velocity field we used in this paper . Recall 
that $u^{there}$ was chosen to obey $u^{there}.dS=0$. 
Dotting \eqref{oe} with  $dS$ we find $A=n_S= n\cdot dS$ from 
which it follows that 
\begin{equation}\label{uoldnew}
u^{there}=\frac{u^{here}-n}{n_S} + dS.
\end{equation}
This is the reason the equations of motion for the uncharged membrane as reported in 
equation 1.7 of \cite{our} apparently do not match with the $Q\rightarrow 0$ limit of the equations of motion we derived in \eqref{relationss} and \eqref{relationsv}. However, we shall see that once we take into account this difference in the definition of $u$,  the uncharged limit of our equations of motion exactly matches with that of \cite{our}.

The equations of motion for the uncharged membrane were reported in 
equation 1.7 of \cite{our} as 
\begin{equation}\begin{split}\label{meour}
& U_{\perp} \cdot K\cdot U_{\perp}   + n_S(n_S^2-1)/S =0, \\
& \left(\left(U_\perp \cdot \nabla\right) u^{there} \right)\cdot P^{\mu\nu}_{there} = 0,\\
& U_\perp= U- (U\cdot n) n, ~~~U= dS+ n_S^2(dS-u^{there}_\mu dx^\mu).
\end{split}
\end{equation}
The projector $P^{\mu\nu}_{there} $ projects orthogonal to the subspace spanned by 
$u^{there}$, $n$ and $dS$. 
But $u_{there}$ is a linear combination of $u_{here}$ and $n$. Therefore
it follows that the projector $P^{\mu\nu}_{there}$ employed in \eqref{meour}
 agree with the projector $P^{\mu\nu}$ in \eqref{relationsv}.
 The covariant derivative `$\nabla$' is a derivative defined in the auxiliary space. In our choice of coordinates, this could be replaces by `$\partial$'.

Using \eqref{uoldnew} we could express the vector $U_{\perp}$ in \eqref{meour}  in terms of 
the velocity $u^{here}$ 
\begin{equation}\label{Uoldnew}
\begin{split}
U_\perp&= U-2n_S n= dS-n_S(u^{here}+n) = n_S(X-u^{here}),\\
\text{where}~~~&X = \frac{dS}{n_S} - n = \left(\frac{n_S}{S} \right)Z.
\end{split}
\end{equation}
Substituting equation\eqref{uoldnew} and \eqref{Uoldnew} in \eqref{meour}, we find \footnote{note that the projected derivative of $\frac{u_{here}-n}{n_S}$ 
equals $\frac{1}{n_S}$ times the projected derivative of $u-n$ as the term 
with $n_S$ differentiated vanishes under projection.
Where $u$ is the new velocity.} 
\begin{equation}\begin{split}\label{menew}
& (X-u^{here}) \cdot K\cdot (X-u^{here})  + \frac{n_S^2-1}{Sn_S }=0 ,\\
& \left[\left(\left( X-u^{here}\right) \cdot \partial\right) (n-u^{here})\right]\cdot P  = 0.\\
\end{split}
\end{equation}
Equations \eqref{menew} exactly match with the $(Q\rightarrow 0)$ limit of  equations \eqref{relationss} and \eqref{relationsv}.

\subsection{Conditions to fix the integration constants}\label{intconst}

As we have explained above, the first order corrections to \eqref{ansatz} 
are obtained by solving a collection of linear ordinary differential 
equations at each point on the membrane. As mentioned above these 
equations turn out to be explicitly solvable and yield regular 
solutions provided the equations of motion of Subsection \ref{eomregh} 
are obeyed. The solutions to these equations are, however, not yet unique. \
as they depend on as yet undetermined integration constants at each membrane 
point. As we have mentioned in the previous subsection, some of these constants 
are determined by the requirement of regularity at $R=0$. This condition 
however leaves several integration constants undetermined. \footnote{As the 
integration `constants' can, in general, be unconstrained functions of 
the membrane world volume (they are constants only in that they do not 
depend on $R$) they are in fact undetermined integration functions on the 
membrane world volume.} In order to obtain a unique solution to our equations 
we will impose additional physically motivated constraints that will 
uniquely determine these integration constants.  \\

{\bf Asymptotic flatness:}\\
An obvious requirement that we impose is that the correction metric and 
gauge field 
$g^{(1)}_{\mu\nu}$ and $A^{(1)}_\mu$ vanish exponentially rapidly as 
$R \to \infty$. This condition ensures that the full spacetime metric 
rapidly approaches the metric of flat space upon moving a large distance 
(in units of $\frac{1}{D}$) away from the membrane. This condition 
sets the value of several integration constants. \\

{\bf Normalization Conditions:}\\

Even after imposing the condition of asymptotic flatness, it turns out that 
we still have two undetermined integration constants in the scalar sector 
and one in the vector sector. This is precisely as should be expected on 
physical grounds. Our starting spacetime \eqref{ansatz} was parameterized by 
two scalar functions (the shape of the membrane and its charge density field) 
plus one vector function (the velocity field). A redefinition of these 
fields (e.g. $Q \rightarrow Q + {\cal O}(1/D)$ leaves \eqref{ansatz} 
unchanged at leading order, but modifies it at first subleading order. 
Such a redefinition will modifies the first order correction to the metric 
by a compensating amount. For this reason we should expect the first order 
correction to have a two parameter ambiguity in the scalar sector and a 
one parameter ambiguity in the vector sector, precisely as we find. 
\footnote{A very similar issue arose in the study of the fluid gravity 
correspondence, and was dealt with in a manner similar to that described 
below. See e.g. \cite{Bhattacharyya:2008jc, 
Bhattacharyya:2008mz, Rangamani:2009xk, Hubeny:2011hd}. }

The ambiguity described above is a result of the fact that we have not
yet supplied a precise all orders definition of the shape, velocity and 
charge density fields that enter into the leading order solution 
\eqref{ansatz}. Such a definition may be supplied by specifying an
additional constraint on all higher order corrections to 
\eqref{ansatz} that would fix the field redefinition ambiguity described 
in the previous paragraph. In this paper we choose to do this by 
requiring that $S_{(VV)}$, $V^{(V)}_{\mu}$ and $S_{(AV)}$ all vanish at 
$R=0$. More invariantly we impose the condition that 
$$ H_{MN} n^N= A_M n^M=0 ~~~{\rm when} ~~\rho=1.$$
We refer to these additional conditions - that effectively define 
the shape, velocity and charge density fields - as `normalization' 
conditions. 

It may be checked (see the upcoming paper \cite{bh}) that the normalization 
conditions we have 
chosen ensure, in particular that the surface $\rho=1$ is a null surface 
which we will later identify with the event horizon of the spacetime. 

The conditions of asymptotic flatness together with 
the normalization conditions are sufficient to fix 
all integration constants, and yield unique expressions for the 
first order correction the the metric and gauge field \eqref{ansatz}.

\subsection{Results for the first order correction on the patch} \label{pertcomp}

In this subsection we present the explicit solution for the 
metric and the gauge field corrections at first order in 
${\cal O}\left(\frac{1}{D}\right)$. Our explicit results 
are presented for ($p=2$), but will be generalized to all $p$ in 
the next section. As mentioned above, our solution takes the form 
\eqref{ukc}. In the rest of this subsection we present our explicit 
results for the functions that appear in \eqref{ukc}\footnote{The solution presented in this subsection depends on three functions $Q$, $S$ and $n_S$. Strictly speaking they should be written as $Q_0$, $S_0$ and $n_S^0$, the values of these functions at $x^\mu =x^\mu_0$. But we did not write it that way firstly because of notational simplicity and secondly because we know that the difference is always suppressed by terms of order  ${\cal O}\left(\frac{1}{D}\right)$.  }

\subsubsection{The functions appearing in the gauge field}
\begin{equation}\label{expansionAi}
\begin{split}
V^A_i(R) =&-\sqrt{2}~Q\left(\frac{S}{n_S}\right)^2\left[(1-Q^2){\mathfrak v}_i^{(5)} + (1+Q^2)\left(\frac{n_S}{S}\right){\mathfrak v}_i^{(2)}\right]Re^{-R}\\
&+\sqrt{2}Q^3\left(\frac{S}{n_S}\right)^2\left({\mathfrak v}_i^{(5)} - \left(\frac{n_S}{S}\right){\mathfrak v}_i^{(2)}\right)\left[1+\log(1-Q^2 e^{-R})\right]e^{-R}\\
\end{split}
\end{equation}

\begin{equation}\label{expansionAX}
\begin{split}
S_{{(Az)}}(R)=
& -\left[\frac{2\sqrt{2} ~S^2Q^3e^{-R}}{(1-n_S^2)(1-Q^2)}\right]\bigg[1+ \log(1-Q^2e^{-R})\bigg]{\mathfrak s}_1\\
&+ \left[\frac{2\sqrt{2}~S^3Q~e^{-R}}{n_S(1-n_S^2)(1-Q^2)}\right]\bigg[\left(Q^2 -R +Q^2R\right)
+Q^2 \log(1-Q^2e^{-R})\bigg]{\mathfrak s}_2.\\
\end{split}
\end{equation}

\begin{equation}\label{expansionAV}
\begin{split}
S_{(AV)}(R) =&~ \sqrt{2}~Q~ R e^{-R}\left(\frac{S}{n_S}\right)\left(\frac{{\mathfrak s}^{(5)}}{Q} -{\mathfrak s}^{(1)} +\frac{S}{n_S}{\mathfrak s}^{(2)}\right)\\
&+2\sqrt{2}\left(\frac{Q^3}{1-Q^2}\right)e^{-R}~\Upsilon_A(R)~\left(\frac{S}{n_S}\right)\left({\mathfrak s}^{(1)} -\frac{S}{n_S}{\mathfrak s}^{(2)}\right) ,\\
\end{split}
\end{equation}\\
where
\begin{equation}
\begin{split}
\Upsilon_A(R) &= \int_0^R dx~\log(1-Q^2e^{-x}).\\
\end{split}
\end{equation}

\subsubsection{The functions appearing in the metric:}

\begin{equation}\label{expansionTensor}
\begin{split}
T_{ij}(R) =\left(\frac{2S}{n_S}\right)\left({\mathfrak t}^{(1)}_{ij} -{\mathfrak t}^{(2)}_{ij} \right)\log(1-Q^2e^{-R}).
\end{split}
\end{equation}

\begin{equation}\label{expansionVX}
\begin{split}
&V^{(z)}_i(R) =\left[\frac{S^2(1+Q^2)}{n_S(1-n_S^2)}\right]\left({\mathfrak v}^{(5)}_i - \left(\frac{n_S}{S}\right){\mathfrak v}^{(2)}_i\right)\log(1-Q^2 e^{-R})
\end{split}
\end{equation}

\begin{equation}\label{expansionVV}
\begin{split}
V^{(V)}_i(R) =&\left(\frac{QS}{n_S}\right)^2\left[1-e^{-R} -f_0(R)\left(1+\log[1-Q^2e^{-R}]\right)\right]\left({\mathfrak v}_i^{(5)} - \left(\frac{n_S}{S}\right){\mathfrak v}_i^{(2)}\right)\\
&-R\left[1-f_0(R)\right]\left(\frac{S}{n_S}\right)^2\left[(1-Q^2){\mathfrak v}_i^{(5)} + (1+Q^2)\left(\frac{n_S}{S}\right){\mathfrak v}_i^{(2)}\right]\\
\end{split}
\end{equation}

\begin{equation}\label{expansionSVX}
\begin{split}
S_{(Vz)}(R) = & S_{(Vz)}^{(1)}(R)~{\mathfrak s}^{(1)}+S_{(Vz)}^{(2)}(R)~{\mathfrak s}^{(2)},\\
S_{(Vz)}^{(1)}(R) =& -\left[\frac{2  Q^2S^2}{(1-n_S^2)(1-Q^2)}\right]\bigg[Q^2\left(e^{-R} -e^{-2R}\right)-f_0(R)~ \log(1-Q^2 e^{-R})\bigg],\\
S_{(Vz)}^{(2)}(R) =&\left[\frac{2Q^2S^3}{n_S(1-n_S^2)(1-Q^2)}\right]\bigg[(e^{-R} -e^{-2R})(Q^2 -R + Q^2R)\\
&~~~~~~~~~~~~~~~~~~~~~~~~~~~~~~~~~-f_0(R)~ \log(1-Q^2e^{-R})\bigg] -\frac{2S^3Re^{-R}}{n_S(1-n_S^2)} .\\
\end{split}
\end{equation}

\begin{equation}\label{expansionSXX}
\begin{split}
&S_{(zz)}(R) =\left[{\mathfrak s}^{(2)}- \left(\frac{n_S}{S}\right)~ {\mathfrak s}^{(1)}\right]\left[\frac{2S^4(1+Q^2)}{(1-n_S^2)^2(1-Q^2)} \right]\log(1-Q^2e^{-R}).\\
\end{split}
\end{equation}

\begin{equation}\label{expansionSTr}
\begin{split}
&S_{(Tr)}(R) =\left[-2  +\left( \frac{S}{n_S}\right)({\mathfrak s}^{(3)}-{\mathfrak s}^{(4)})\right]\log(1-Q^2e^{-R}).\\
\end{split}
\end{equation}

\begin{equation}\label{expansionSVV}
\begin{split}
S_{(VV)}(R)=&~-\sqrt{2} Q~ e^{-R}S_{AV}(R)  +Q^2\left[e^{-2R} - e^{-R}\right]\\
& +2e^{-R}\left[Q^2~R ~\left(\frac{{\mathfrak s}^{(5)}}{Q} -{\mathfrak s}^{(1)} +\frac{S}{n_S}{\mathfrak s}^{(2)}\right)
+\Upsilon_H(R)\left({\mathfrak s}^{(1)} -\frac{S}{n_S}{\mathfrak s}^{(2)}\right)\right] ,\\
\end{split}
\end{equation}\\
where
\begin{equation}
\begin{split}
f_0(R) &= 1-\left[(1+Q^2)e^{-R} -Q^2e^{-2R}\right],\\
\Upsilon_H(R)&=\left[(e^{-R}-Q^2)\log(1-Q^2e^{-R})-(1-Q^2)\log(1-Q^2) + \left(\frac{Q^2(1+Q^2)}{1-Q^2}\right)\Upsilon_A(R)\right].
\end{split}
\end{equation}

\subsubsection{Correction (2nd order in $\frac{1}{D}$) to the scalar field $\phi$}
\begin{equation}\label{expansionPhi}
\begin{split}
&\delta\phi = \sum_{k=1}\left(\frac{1}{D}\right)^k \delta\phi^{(k)},\\
&\delta\phi^{(1)} =0,\\
&\delta\phi^{(2)}(R) =-2 S_{(Tr)}(R) - \left(\frac{1-n_S^2}{S^2}\right)S_{(zz)}(R).\\
\end{split}
\end{equation}\\

\subsubsection{ The $Q\rightarrow 0$ limit}
If we set $Q$ to zero in equation \eqref{expansionAi} to \eqref{expansionPhi},
 most of the functions vanish except $V^{(V)}_i$ and $S_{(Vz)}$. In the uncharged limit, the metric takes the following simple form,
\begin{equation}\label{unchargeda}
\begin{split}
G_{ab}^{(1)} dq^a dq^b|_\text{uncharged} =-2Re^{-R}\bigg[\left(\frac{S}{n_S}\right)^2\left({\mathfrak v}^{(5)}_i+ \frac{n_S}{S} {\mathfrak v}^{(1)}_i  \right)dy^i +\frac{S^3~{\mathfrak s}^{(2)}}{n_S(1-n_S^2)} dz\bigg]dV.
\end{split}
\end{equation}

\subsection{The global first order metric}\label{corrpm1}

With the first order corrected patch metric in hand (see the previous 
subsection), it is straightforward
to find the global form of the metric and gauge field which, when expanded
in any patch around a membrane point, will reproduce the results 
of Appendix \ref{pertcomp}. In order to obtain this global form 
we simply make the replacements 
$$e^R \rightarrow \rho^{-D}, ~~~R \rightarrow D\times(\rho -1), ~~~
dV  \rightarrow O_Mdx^M, ~~~dR \rightarrow D\times(d \rho),$$
in the results of subsection \ref{pertcomp}. The final metric 
obtained in this manner is already reasonably compact. There is, 
however, a physically motivated rewriting of this result in a form 
that is both more elegant and also makes manifest the `geometrical'
nature of our final result, as we explain in more detail in the next 
section. 

\section{Geometrical Form of the first order corrected metric}\label{efgf}

\subsection{Redistribution invariance and the Geometrical form}\label{redinv}

The membrane equations \eqref{relationss} and \eqref{relationsv} make 
make special reference to $e^\phi$, $n_S$ and the one-form field 
$Z_\mu$. The same is true of our explicit results for the first order 
correction to \eqref{ansatz}, presented in subsection \ref{pertcomp}.
Expressions involving $S$, $n_S$ and $Z$ of course are only well 
defined for configurations that preserve an SO$(D-p-2)$ symmetry. Moreover
the definition of, e.g. $S$ depends on the details of the isometry.

Unconstrained dependence on $S$ and $n_S$ is unacceptable for the following
reason. A solution that preserves an SO$(D-p-2)$ isometry also preserves 
an SO$(D-p'-2)$ isometry for all $p'>p$. It follows that any solution 
of the equations for a particular choice of $p$ must also be a solution 
of the same equation for all larger $p$. We refer to this requirement 
as the requirement of redistribution invariance.

The requirement of redistribution invariance is most simply met if the 
equation of motion and the metric and gauge field can both be written in an 
explicitly geometrical form that makes no reference to the particular 
isometry group of the solution. The membrane equation and first order metric 
and gauge field obtained in this section do indeed turn out to have this 
property. 

The reader may, at first, wonder how it is possible for expressions 
with explicit appearances of $S$ and $n_S$ to also be geometrical. This 
is, infarct, possible in the large $D$ limit, as we now explain with an 
example. Consider the manifestly geometrical expression 
$\nabla^2 \rho$ where $\nabla$ refers to the covariant derivative on the 
full flat $D$ dimensional embedding spacetime. Let us now evaluate this 
expression in the large
$D$ limit restricting attention to membrane configurations that preserve 
an SO$(D-p-2)$ isometry. The computation is most conveniently performed 
using the following coordinates  
$$ds^2=\eta_{\alpha\beta} dw^\alpha dw^\beta + dS^2 + S^2 d \Omega_d^2,$$
in the embedding flat space. Using these coordinates 
$$\nabla^2 \rho= \frac{1}{S^d}\partial_\mu \left( S^d \partial^\mu \rho \right).$$
At leading order in the large $D$ limit this expression reduces to 
$D \frac{ dS\cdot  d \rho}{S}$
It follows that $\frac{ dS\cdot  d \rho}{S}  =\frac{\nabla^2 \rho}{D}$ at
leading order in the large $D$ limit. Consequently any appearance 
of  $\frac{ dS\cdot  d \rho}{S}$ in any equation may be explicitly geometrized. 

Similar manipulations allow us to geometrize several other expressions 
involving $S$, $n_S$ and $Z$. Of course not every expression involving these 
quantities can be geometrized (expressions that are not redistribution 
invariant certainly cannot). However it turns out that all terms in 
the equations of motion \eqref{relationss} and \eqref{relationsv} and all 
terms in our explicit expression for the metric and gauge field in 
subsection \ref{pertcomp} can be geometrized. The final geometrical 
expressions for equations of motion and the the first order corrected metric 
and gauge field are more compact than the unprocessed expressions.
In the next section we present our final results for the first order 
corrections to the metric and gauge field in explicitly geometrical form. 
In the subsequent subsection we do the same for the equations of motion.

\subsection{Metric and Gauge field in Geometric Form} \label{mggf}

While we expect the first order correction to the metric and gauge field 
to be geometrizable on physical grounds, this requirement is nontrivial 
at the algebraic level. The vector $Z_\mu$ - which is treated as a special 
in the computation described above and in subsection \ref{pertcomp}- has 
no intrinsic geometrical significance \footnote{On the other hand the 
vectors $n^\mu$ and $u^\mu$ are intrinsically geometrical, as they describe
the membrane shape and velocity field in $D$ dimensional spacetime.}.
If the first order correction to the metric and gauge field is completely 
geometrical, it should be possible to rewrite it in a manner that makes 
no reference to $Z_\mu$. In fact it should be possible to rewrite the 
metric and gauge field in the form
\begin{equation}\begin{split}\label{metgform}
h_{MN} &= F(\rho) O_M O_N +H^{(T)}_{MN} + 2O_{(M} H_{N)}^{(V)} + H^{(S)} O_M O_N + H^{(Tr)} {\cal P}_{MN},\\
A_M&= \sqrt{2} Q ~\rho^{-(D-3)}~ O_M +\left( A^{(S)}O_M + A^{(V)}_M\right),\\
\text{where}&\\
F(\rho)&=\left[(1+Q^2)\rho^{-(D-3)} -Q^2 \rho^{-2(D-3)}\right],\\
{\cal P}_{MN}&= \eta_{MN}-O_M n_N -O_N n_M + O_M O_N,\\
{\cal P}^{MN}& H_{N}^{(V)}={\cal P}^{MN} A_{N}^{(V)}=0, ~~~{\cal P}^{MN} H^{(T)}_{MQ} =0, ~~~{\cal P}^{MN}H^{(T)}_{MN}=0,
\end{split}
\end{equation}
\eqref{metgform} should reproduce the expressions for $g_{\mu\nu}$, 
$A_\mu$ (see \eqref{uk}) as well as the scalar $\phi$ (recall that 
$\phi$ is part of the full $D$ dimensional metric). 

The general metric and gauge field presented in \eqref{metgform} 
are parameterized by three unknown scalar functions (rather than the seven scalar functions in \eqref{uk} and in the 
expansion of the scalar $\phi$) and by two vector functions 
(rather than three vector functions in \eqref{uk}). 
It follows that the explicit results of subsection \ref{pertcomp} can be 
recast into the form \eqref{metgform} only if the seven scalar functions
determined in subsection \ref{pertcomp} obey four constraints, and the 
three vector functions determined in the same Appendix obey a single 
constraint equation.  

We have verified that our explicit results do infarct obey all constraints. 
We view this fact as an impressive consistency check of the complicated algebra
that went into obtaining the explicit results of subsection \ref{pertcomp}.

\begin{table}[ht]
\vspace{0.5cm}
\centering 
\begin{tabular}{|c| c| }
\hline
 &${\mathfrak S}_{(1)} = \left( \frac{D}{{\mathcal K}}\right)\left[\frac{u\cdot\partial Q}{Q} -u\cdot K\cdot u + \frac{(u\cdot\partial){\mathcal K}}{{\mathcal K}}\right]$ \\
Scalars&\\
&${\mathfrak S}_{(2)}= \left( \frac{D}{{\mathcal K}}\right)\left[ u\cdot K\cdot u - \frac{(u\cdot\partial){\mathcal K}}{{\mathcal K}}\right]$\\
\hline
&$ {\mathfrak V}_{(1)}^M = \left( \frac{D}{{\mathcal K}}\right)\left[\frac{\nabla_N {\mathcal K}} {\mathcal K} + (u\cdot\nabla)u_N\right]~{\cal P}^{NM}$ \\
Vectors&\\
&${\mathfrak V}_{(2)}^M=  \left( \frac{D}{{\mathcal K}}\right)\left[\frac{\nabla_N {\mathcal K}} {{\mathcal K}} - (u\cdot\nabla)u_N\right]~{\cal P}^{NM}$ \\
\hline
&\\
Tensor& ${\mathfrak T}^{MN} =  {\cal P}^{MQ_1}~\left(\frac{D}{{\cal K}}\right)\left[\frac{\nabla_{Q_1} O_{Q_2}+\nabla_{Q_2} O_{Q_1}}{2}-\eta_{Q_1 Q_2}\left(\frac{\nabla\cdot O}{D-2}\right) \right]~{\cal P}^{Q_2N}$\\
\hline
\end{tabular}\vspace{.5cm}
\caption{We list the data that enters into our 
explicit results for the first order correction to the metric and the 
gauge field. All data is presented in explicitly geometrical form.
 $\rho$, $Q$ and $u^\mu$ should be thought of as two functions 
and a vector field in flat $D$ dimensional space. All derivatives that 
appear in this table are covariant derivatives w.r.t flat $D$ dimensional 
space. }
\label{table:geometric} 
\end{table}
\noindent

As our explicit results obey all consistency conditions, it is possible to 
rewrite our final results in the explicitly geometric form \eqref{metgform}.  
We find that the various free functions in \eqref{metform} are given by  
\begin{equation}\label{gaugegeo}
\begin{split}
A^{(V)}_M=&~-\left(\frac{\sqrt{2}}{D}\right)Q\rho^{-D}\bigg[D(\rho-1)({\mathfrak V}_{(1)}-Q^2~{\mathfrak V}_{(2)}) - Q^2 [1+\log(1-\rho^{-D}Q^2)]{\mathfrak V}_{(2)}\bigg]_M\\
& + {\cal O}\left(\frac{1}{D}\right)^2,\\
A^{(S)} =&~\left(\frac{1}{D}\right)\bigg[ \sqrt{2}~Q~D (\rho-1)~\rho^{-D} {\mathfrak S}_{(1)} + 2\sqrt{2}\left(\frac{Q^3}{1-Q^2}\right)\rho^{-D}~\Upsilon_A(\rho)~{\mathfrak S}_{(2)}\bigg] + {\cal O}\left(\frac{1}{D}\right)^2.\\
\end{split}
\end{equation}
\begin{equation}\label{metricgeo1}
\begin{split}
H^{(T)}_{MN}=&~\left(\frac{2}{D}\right)\log(1-Q^2\rho^{-D})~{\mathfrak T}_{MN} + {\cal O}\left(\frac{1}{D}\right)^2,\\
H^{(V)}_M =&~
\left(\frac{1}{D}\right)\bigg\{Q^2\left[(F(\rho)-\rho^{-(D-3)}) +(F(\rho)-1)\log(1-Q^2\rho^{-D})\right]{\mathfrak V}_{(2)M}\\
&-D(\rho-1) F(\rho)~[{\mathfrak V}_{(1)}-Q^2~{\mathfrak V}_{(2)}]_M \bigg\}+ {\cal O}\left(\frac{1}{D}\right)^2.\\
\end{split}
\end{equation}
\begin{equation}\label{metricgeo2}
\begin{split}
H^{(S)}=&~-\sqrt{2} Q~ \rho^{-D}A^{(S)} +\left(\frac{1}{D}\right)\bigg[\rho^{-(D-3)} - F(\rho)\bigg]\\
& + \left(\frac{2}{D}\right)\rho^{-D}\left[Q^2~ D(\rho-1) ~ {\mathfrak S}_{(1)}
+\Upsilon_H(\rho){\mathfrak S}_{(2)}\right] +{\cal O}\left(\frac{1}{D}\right)^2,\\
\\
H^{(Tr)}=& ~ {\cal O}\left(\frac{1}{D}\right)^3,
\end{split}
\end{equation}
where 
\begin{equation} \label{fdef}
\begin{split}
F(\rho) &= \left[(1+Q^2)\rho^{-(D-3)} -Q^2 \rho^{-2(D-3)}\right],\\
\Upsilon_A(\rho) &= \int_0^{D(\rho-1)} dx~\log(1-Q^2e^{-x}),\\
\Upsilon_H(\rho)&=\left[(\rho^D-Q^2)\log(1-Q^2\rho^{-D})-(1-Q^2)\log(1-Q^2) + Q^2\left(\frac{1+Q^2}{1-Q^2}\right)\Upsilon_A(\rho)\right].
\end{split}
\end{equation}

\subsubsection{The limit $Q \rightarrow 0$} 

The results of the previous subsection simplifies drastically 
in the limit $Q \to 0$. In this limit the gauge field simply vanishes, and 
the full first order corrected 
 metric is given by the remarkably simple expression
\begin{equation}\label{unchargedgeo}
\begin{split}
ds^2_{uncharged} =& ds_\text{flat}^2 + \rho^{-(D-3)} (O_M dx^M)^2\\
& -2(\rho-1)\rho^{-(D-3)} [{\mathfrak V}_{(1)}]_M O_N dx^M dx^N
+ {\cal O}\left(\frac{1}{D}\right)^2,
\end{split}
\end{equation}
where ${\mathfrak V}_1$ is defined in table \ref{table:geometric}.

In Appendix \ref{App:compare} we have shown how this geometric form of the metric and gauge field reduce to the solution presented in subsection \ref{pertcomp}, once we impose the constraint of SO$(D-p-2)$ invariance on 
all geometric data.

\subsection{Geometrizable form of the membrane equations of motion}\label{eomgf}

The membrane equations of motion \eqref{relationss} and \eqref{relationsv} 
may be recast into a simpler looking form.  We have a combined equation 
capturing both vector equation and one of the scalar equations. 
\begin{equation}\label{chdeqn1}
\bigg[(u-X)\cdot{\tilde \nabla} O  - Q^2 (u\cdot{\tilde \nabla})u+Q^2(X\cdot K) \bigg]\cdot{\cal P} +\left(\frac{n_S}{S}\right)(1-Q^2)X  = 0,
\end{equation}

\begin{equation}\label{chdeqn2}
\begin{split}
&(X-u)\cdot{\tilde \nabla} Q + Q\left[\left(\frac{S}{n_S}\right)(u\cdot{\tilde \nabla})\left(\frac{n_S}{S}\right) -(u\cdot K\cdot u)\right]=0,\\
\end{split}
\end{equation}
where
\begin{equation*}
\begin{split}
&{\cal P}_{\mu\nu} =\text{Projector perpendicular to $u_\mu$ and $n_\mu
$},\\
& X = \frac{dS}{n_S} -n =  \left(\frac{S}{n_S} \right)Z,~~~~
O= n-u.
\end{split}
\end{equation*}
In this equation ${\tilde \nabla}$ above is the partial derivative on 
the membrane world volume viewed as a submanifold of \eqref{auxspace}.

The projection of  equation \eqref{chdeqn1} perpendicular to $Z_\mu$ 
directly reduces to the vector equation of motion as given in equation 
\eqref{relationsv}. In appendix \ref{app:eqnrel} we have shown that equation 
\eqref{chdeqn2} is equal to second equation of \eqref{relationss}.  Moreover 
the dot product \eqref{chdeqn1} with $Z_\mu$ equals the first equation in 
\eqref{relationss} upto correction of ${\cal O}\left(\frac{1}{D}\right) $.

We re emphasize that the projector employed in \eqref{chdeqn1} projects 
orthogonal to $n$ and $u$ but not to $Z_\mu$. In other words  \eqref{chdeqn1}
unifies a SO$(p)$ scalar and SO$(p)$ vector equation into a single 
`geometrical' vector equation. This fact may lead the reader to suspect that 
the equations \eqref{chdeqn1} and \eqref{chdeqn2} are geometrizable (i.e. 
can be written without any explicit reference to the isometry direction. 
This is indeed the case. It is not too difficult to demonstrate that 
the geometric form of the equations of motion, \eqref{Memeqsc} (see the 
introduction) reduce immediately to \eqref{chdeqn1} and \eqref{chdeqn2} 
upon using the dictionary of translation as presented in appendix \ref{translation}.

\subsubsection{Constraint equations and the membrane equations of motion}

In the previous section we explained that the Einstein and Maxwell constraint
equations play a special role in our construction. We obtained the 
membrane equations of motion from the requirement that these bulk equations 
admit nonsingular solutions. Once the membrane equations were imposed, 
it was possible to utilize the constraint equations to solve for some 
unknown bulk scalar and vector fields in terms of others in a nonsingular
manner. We have already mentioned in subsection \ref{eomregh} that the 
geometric nature of the membrane equations is a direct consequence of 
the geometric nature of Einstein's constraint equations.

In this subsection we wish to focus on the fact the constraint equations 
played two roles in the perturbative program of the previous section.
\begin{itemize}
\item{1} They yielded the membrane equations of motion.
\item{2} They allowed us to solve for some unknowns bulk fields 
in terms of others. 
\end{itemize}

Interestingly enough, the relations that we obtain from item (2) above are 
all {\it automatic} in the expression \eqref{metgform}. In other words 
the relations of item (2) above are simply a subset of the relations 
between various unknown bulk vectors fields and various 
 unknown bulk scalar scalars fields that are forced on us once we assume 
that the first order metric and gauge field correction take on a 
geometric form. 

Had we used hindsight to set up our perturbative expansion in a
manifestly geometric manner by simply assuming that our first order 
correction takes the form \eqref{metgform} then the constraint equations 
of subsection \ref{eomregh} would simply have {\it reduced} to the 
membrane equations \eqref{Memeqsc}, exactly as in the studies of 
the fluid gravity correspondence (see e.g. \cite{Bhattacharyya:2008jc, Erdmenger:2008rm, Banerjee:2008th, Bhattacharyya:2008mz, 
Rangamani:2009xk, Hubeny:2011hd}).

Recall that the Einstein constraint equations assert the conservation of 
the Brown York Stress tensor, while the Maxwell constraint equation asserts
the conservation of a `charge current' $F^{\rho M}$. These observations 
suggest that it may be possible to recast our membrane equations 
\eqref{Memeqsc} as conservation equations for a manifestly geometric 
membrane stress tensor and charge current, as was the case in the 
study of fluid gravity.  We will not pursue this point further in this paper 
but hope to return to it in the near future.

\subsection{Comparison with the Reissner-Nordstrom solution} \label{rns}

As an elementary check of the results reported in Subsection \ref{mggf}, 
consider the following membrane configuration. Let 
$u_M dx^M= -dt$, $Q = const$ and let the membrane 
surface be given by $x^M(\eta_{MN} + u_M u_N) x^N=r_0^2$. 
It is easily verified that this configuration solves the membrane 
equations \eqref{Memeqsc}; clearly this static soap bubble 
solution is dual to the Reissner-Nordstrom black hole 
\eqref{bhgfkb}.

We will now use the formalism developed in this paper to determine 
the spacetime metric and gauge field dual to this membrane
solution, to first order in $1/D$. 

Let us first start with the leading order solution \eqref{ansatz} dual 
to this solution. We need to find a function $\rho$ that obeys the first of equation 
\eqref{properties} and s.t. $\rho=1$ on the membrane surface listed above. 
The unique solution to this mathematical problem is given by 
$$\rho=\sqrt{ \frac{x^M(\eta_{MN} + u_M u_N) x^N}{r_0^2}}.$$
Next we must determine the spacetime fields $u$ and $Q$ fields 
that reduce to $-dt$ and $q$ on the membrane and obey 
\eqref{properties} everywhere in the bulk of $D$ dimensional 
flat space. The unique solution to this problem is given by 
$u=-dt$ and $Q$ in all of flat space. The leading order 
solution with this data is given by 
\eqref{ansatz} with these choices for $\rho$, $Q$ and $u$. 

Let us now turn to the first order correction. It is easily 
verified that relevant geometric data as given in table \ref{table:geometric} vanish on this particular profile of $\rho$ and $u_M$ 
and $Q$. It follows that the first order correction to the 
gauge field vanishes. The first order correction to the metric
also almost vanishes. Of all the quantities listed in \eqref{gaugegeo}, 
\eqref{metricgeo1} and \eqref{metricgeo2}
vanish except for $H^S$ which evaluates to 
$\left(\frac{1}{D}\right)\bigg[\rho^{-(D-3)} - F(\rho)\bigg]$ 
with $F(\rho)$ listed in \eqref{fdef}. 
 
Plugging these values of $\rho$, $u_M$ and $Q$ into  \eqref{ansatz} 
and adding the correction terms \eqref{metgform} , it 
 follows that the metric and gauge field dual to our simple solution of the 
membrane equations is given, to first order in $\frac{1}{D}$, by
\begin{equation} \label{metgf}
\begin{split}
g_{MN} &=\eta_{MN} + \left[(1+Q^2) \rho^{-(D-3)} - Q^2\rho^{-2(D-3)}\right]  O_M O_N\\
&+\frac{Q^2}{D}\left(\rho^{-2(D-3)} - \rho^{-(D-3)}\right) O_M O_N + {\cal O}\left(\frac{1}{D}\right)^2,
\end{split}
\end{equation}

\eqref{metgf} is easily seen to agree with the exact RN black 
hole solution \eqref{bhgfkb} expanded to leading nontrivial 
order in $1/D$. The function $\rho$ of \eqref{bhgfkb} agrees exactly 
with the function $\rho$ reported above. 
The only appearance of $D$ in the solution
\eqref{bhgfkb} is in the factor $c_D$. Upon plugging the expansion 
$$c_D= 1 - \frac{1}{D} + {\cal O}\left(\frac{1}{D}\right)^2,$$
into \eqref{bhgfkb} we immediately recover \eqref{metgf}.

The matching performed in this subsection was almost trivial. 
In the upcoming paper \cite{bh} we use the same method to 
match the metric dual to a rigidly rotating 
solution of the membrane equations to the much more complicated
exact solution of an uncharged rotating Myers Perry black hole \cite{Myers:1986un}. 
Once again we find a perfect match between the two expansions.

\section{Light quasinormal spectrum or the RN black hole} \label{qnm}

Our membrane equations \eqref{Memeqsc} should describe all SO$(D-p-2)$ 
invariant black hole dynamics (over times scales much larger than 
$1/D$) at large $D$. As a first application of these 
equations, in this section 
we will use them to obtain a prediction for the spectrum of light 
quasinormal modes (those with frequencies of order unity rather than 
of order $D$) about charged Reissner-Nordstrom black holes in 
the large $D$ limit. 

For the purposes of this section we work in polar coordinates in $D$ 
spacetime dimensions. Our coordinate system for flat space is given by 
\begin{equation}\label{fspc}
ds^2= -dt^2+ dr^2 + r^2 d \Omega_{D-2}^2~~.
\end{equation}
The exact solution of \eqref{Memeqsc} dual to RN black holes was 
presented in subsection \ref{rns}. In the coordinates \eqref{fspc} this 
solution takes the particularly simple form 
\begin{equation} \label{bhpc}
r=1, ~~~Q=Q_0={\rm const}, ~~~u=-dt,
\end{equation}
where we have chosen units that set the size of the membrane to unity. 
\footnote{We do not loose generality by making this choice. The classical 
Einstein Maxwell equations studied in this paper enjoy 
invariance under the following `scaling' symmetry: 
$${\tilde g}_{MN}=\alpha^2 g_{MN}, ~~~{\tilde F}_{MN} =\alpha F_{MN}.$$ 
This scale transformation together with the coordinate change 
${\tilde x}^M  = \alpha x^M$ transforms a Reissner Nordstrom black hole with 
Schwarzschild radius $r_0$ and charge parameter $Q_0$ into a Reissner Nordstrom 
black hole with Schwarzschild radius $\alpha r_0$ and charge parameter $Q_0$. 
It follows that the quasinormal mode frequencies of the black hole
parameterized by $(r_0, Q_0)$ are simply $\frac{1}{r_0}$ times those 
for the black hole parameterized by $(1, Q_0)$. For this reason we 
will perform all computations in this section with black holes of radius 
unity, and simply reinsert factors of $r_0$ in the final answer. }

The most general linearized perturbation around \eqref{bhpc} takes the form 
\begin{equation}\label{pert}
\begin{split}
r &= 1 + \epsilon~\delta r(t,\theta),\\
Q &= Q_0 + \epsilon~\delta Q(t,\theta),\\
u &= -dt + \epsilon~\delta u_{\mu}(t,\theta)dx^{\mu}.
\end{split}
\end{equation}
We now adopt the following strategy. We 
simply insert the expansions \eqref{pert} into 
\eqref{Memeqsc}, work to linear order in $\epsilon$ and obtain the 
effective linear equations for the fluctuation fields $\delta r$, 
$ \delta Q$ and $\delta u_\mu$ 
defined in \eqref{pert}. These fields live on the membrane world volume. 
A useful set of coordinates on this world volume are the angular coordinates
$\theta^a$ on $\Omega_{D-2}$ and time. The metric on the membrane world 
volume in these coordinates is obtained by inserting the first of 
\eqref{pert} into \eqref{fspc}, and is given in terms of the function 
$\delta r(t, \theta)$. To linear order in $\epsilon$ the metric on the membrane 
surface is given by 
\begin{equation}\label{metonmemb}
ds^2= -dt^2 + \left(1+2 \epsilon \delta r \right)d\Omega_{D-2}^2~~.
\end{equation}

It is useful to have a dictionary to go between forms and vectors 
that live on the membrane and those that live in spacetime. Consider 
a vector field $A^\mu$ that lives on the membrane. This vector field may 
be uplifted to spacetime. The spacetime components $A^M_{(ST)}$ of this 
vector field are given in terms of the world volume components $A^a$ by 
the formulae 
\begin{equation} \label{stmem}
A^a_{(ST)}=A^a, ~~~A^t_{(ST)}=A^t, ~~~A^r_{(ST)}= \epsilon \left( 
A^t\partial_t \delta r + A^a\partial_a \delta r  \right).
\end{equation}
In a similar manner, a one-form field in spacetime $B^{(ST)}_a$ is easily pulled 
back to a one-form field $B_a$ on the membrane. In formulae 
\begin{equation} \label{stmemf}
B_a=B_a^{(ST)} +\epsilon B_r^{(ST)}\partial_a \delta r , ~~~
B_t=B_t^{(ST)} +\epsilon B_r^{(ST)}\partial_t \delta r .
\end{equation}
As a simple consistency check on these formulae, it is easily verified that 
$A^\mu B_\mu= A^M_{(ST)} B_M^{(ST)}$. 
Below we will treat the field $u_\mu$ in \eqref{pert} as a one-form field 
on the membrane. Recall that $u_\mu$ is constrained by the requirement 
$\nabla\cdot u=0$, i.e. that the velocity field is divergence free. 
Here $\nabla$ is the covariant derivative taken in the metric 
\eqref{metonmemb}. 

In order to evaluate the terms in \eqref{Memeqsc} we need to compute the, 
normal one-form and extrinsic curvature of the membrane as well as a few derivatives
of the velocity field. The computations involved are straightforward: 
to linear order in $\epsilon$ we find 
\begin{equation} \label{componmem}
\begin{split}
n_r &= 1,\\
n_{\mu} &= -\epsilon\partial_{\mu}\delta r,\\
K_{tt} &= -\epsilon\partial_t^2\delta r,\\
K_{ta} &= -\epsilon\partial_t\nabla_a\delta r,\\
K_{ab} &= -\epsilon\nabla_a\nabla_b\delta r + (1 + \epsilon\delta r)g_{ab},\\
\delta u_t &= 0, ~~~~~~~~~~(u\cdot u =-1)\\
(u\cdot K)_t &= K_{tt} = -\epsilon\partial_t^2\delta r,\\
(u\cdot K)_a &= -\epsilon\partial_t\nabla_a\delta r + \epsilon\delta u_a,\\
\mathcal{K} &= {K_A}^A = D\left(1-\epsilon\left(1+\frac{\nabla^2}{D}\right)\delta r\right),
\end{split}
\end{equation}
where $a,b$ are angular directions on $\Omega_{D-2}$, the symbol $\mu$ 
runs over these angular coordinates and time (i.e $\mu=(t,a)$) 
and $g_{ab}$ is the round metric on $S^{D-2}$. All indices 
in \eqref{componmem} are indices on the spherical metric world volume, 
i.e. on a space with metric
\begin{equation}\label{auxmet}
ds^2=-dt^2+ d\Omega_{D-2}^2~~,
\end{equation}
and all covariant derivatives in \eqref{componmem} are 
taken in the background spacetime \eqref{auxmet}.

Using \eqref{componmem}, the first equation in \eqref{Memeqsc} may be 
shown to reduce, at linearized order in $\epsilon$, to 
\begin{equation}\label{linun}
\left(1+\frac{\nabla^2}{D}\right)\delta u_a + (1-Q_0^2)\nabla_a\left(1+\frac{\nabla^2}{D}\right)\delta r-\partial_t\nabla_a\delta r -(1+Q_0^2) \partial_t\delta u_a=0.
\end{equation}
(All covariant derivatives are once again evaluated on the metric 
\eqref{auxmet}). 

As we have noted above, the fluctuation velocity field $\delta u_a$ is 
constrained by the condition $\nabla\cdot u=0$. The divergence in this equation
is evaluated in the full membrane metric. Rewriting this constraint 
in terms of fields that are taken to propagate on the fixed 
metric \eqref{auxmet} (to linear order in $\epsilon$ and leading order
in $D$) we find 
\begin{equation}\label{vaux}
\nabla_\mu \delta u^\mu = - D \partial_t \delta r,
\end{equation}
with the covariant derivatives now evaluated on the metric \eqref{auxmet}. 
From now on until the end of this section our fluctuation fields will all 
be taken to propagate on the fixed background \eqref{auxmet} and all 
covariant derivatives will refer to this metric 
 unless explicitly otherwise declared. 

As $\delta u^t$  vanishes (this follows upon linearizing the equation 
$u\cdot u=-1$), 
\eqref{vaux} may be rewritten as 
\begin{equation}\label{vauxa}
\nabla_a \delta u^a = - D \partial_t \delta r.
\end{equation}
In order to solve this equation it is useful to define 
\begin{equation}\label{udecomp}
\delta u_a = \nabla_a \Phi + \delta v_a,
\end{equation} 
where
\begin{equation}\label{divfree}
\nabla\cdot \delta v=0.
\end{equation}
It follows from \eqref{vauxa} that 
\begin{equation} \label{phconst}
\nabla^2 \Phi= -D \partial_t \delta r.
\end{equation}
Below we will use this equation to eliminate $\Phi$ in favour of $\delta r$. 
Note that \eqref{phconst} admits a solution if and only if its RHS has 
no overlap with the kernel of the operator $\nabla^2$. As the kernel of 
$\nabla^2$ consists of functions that are constant on the sphere, it follows 
that \eqref{phconst}  is consistent if 
and only if the spatially constant (i.e. $l=0$ mode)  of $\delta r$ is 
time independent. When this condition is obeyed, $\Phi$ may be solved for 
in terms of $\delta r$, as we will do below. 

Plugging the expansion \eqref{udecomp} into the equation
\eqref{linun} we find 
\begin{equation} \label{vecsc}
\begin{split}
\left(1+\frac{\nabla^2}{D}-(1+Q_0^2) \partial_t\right)\delta v_a =& - \left((1-Q_0^2)\nabla_a\left(1+\frac{\nabla^2}{D}\right)-\partial_t\nabla_a\right)\delta r\\ &-\left(1+\frac{\nabla^2}{D}-(1+Q_0^2) \partial_t\right)\nabla_a\Phi.
\end{split}
\end{equation}

\subsection{The spectrum of shape fluctuations}\label{spshape}

Taking the divergence of \eqref{vecsc} and using \eqref{divfree} and 
 \eqref{phconst} 
we obtain the following decoupled scalar equation for the fluctuation 
field $\delta r$ 
\begin{equation}\label{decsc}
D(1+Q_0^2)\partial_t^2\delta r - 2D\left(1+\frac{\nabla^2}{D}\right)\partial_t\delta r+(1-Q_0^2)\left(1+\frac{\nabla^2}{D}\right)\nabla^2\delta r =0.
\end{equation}
\footnote{In order to obtain \eqref{decsc} we have used
and
\begin{equation}\label{gradlaplcomm}
\begin{split}
\nabla^a\nabla^2\delta u_a &= \nabla^2\nabla_a\delta u^a + R^{ab}\nabla_a\delta u_b,\\
&= \nabla^2\nabla_a\delta u^a + D~g^{ab}\nabla_a\delta u_b,\\
&= D\left(1+\frac{\nabla^2}{D}\right)\nabla_a\delta u^a = -D^2\left(1+\frac{\nabla^2}{D}\right) \partial_t \delta r.
\end{split}
\end{equation} }

The most general linearized fluctuation $\delta r$ can be expanded as 
\begin{equation}\label{drexp}
\delta r= \sum_{l,m}a_{lm} Y_{lm} e^{-i \omega^r_l t}~~.
\end{equation} 
Here $Y_{lm}$ are spherical harmonics on $S^{D-2}$, $l$ labels the spherical 
harmonic representation,  $m$ is a collective label for all the internal 
quantum numbers within a given spherical harmonic representation. 

Let us pause to give a more complete description of scalar spherical harmonics
in arbitrary dimensions, and in particular to compute the eigenvalue 
under $\nabla^2$ acting on the $l^{th}$ spherical harmonic. The $l^{th}$ 
spherical harmonic, $Y_{lm}$, are composed of the collection of functions 
on $S^{D-2}$ obtained by 
restricting homogeneous degree $l$ polynomials in $R^{D-1}$ to the unit sphere.
The polynomials in questions are linear combinations  of monomials of the form 
$a_{\mu_1 \mu_2 \mu_3 \ldots \mu_l} x^{\mu_1} x^{\mu_2} \ldots x^{\mu_l}$ where 
$a_{\mu_1 \mu_2 \mu_3 \ldots \mu_l}$ are symmetric and traceless tensors. 
It is easily shown that 
\begin{equation} \label{sphericalharm}
-\nabla_{S^{D-2}}^2 Y_{lm} = l(D+l-3) Y_{lm}.
\end{equation} 
\footnote{This may be demonstrated as follows. The condition of 
tracelessness ensures that the degree $l$ polynomials described above obey the 
equation $\nabla^2 \Phi=0$, where $\nabla^2$ is evaluated in $R^{D-1}$. But 
\begin{equation}\label{lapusp}
0=\nabla_{R^{D-1}}^2 \Phi= \frac{1}{r^{D-2}} \partial_r 
\left(r^{D-2}\partial_r r^l \right) + \frac{\nabla^2_{S^{D-2}}\Phi}{r^2}.
\end{equation}
(the RHS of this equation is $\nabla^2$ of the function in $R^{D-1}$ 
evaluated in polar coordinates). Here 
 $\nabla^2_{S^{D-2}}$ is the Laplacian evaluated 
on the unit sphere. \eqref{sphericalharm} follows from \eqref{lapusp}.}

Plugging the expansion \eqref{drexp} into \eqref{decsc} and using 
\eqref{sphericalharm} we find (at leading order in large $D$) 
\begin{equation}\label{scsp}
\omega_l^r = \frac{-i(l-1)\pm\sqrt{(l-1)(1-lQ_0^4)}}{1+Q_0^2}.
\end{equation} 
Re inserting factors of $r_0$ (see the discussion in the introduction 
to this section) we find 
\eqref{sphericalharm} we find (at leading order in large $D$) 
\begin{equation}\label{scspf}
r_0 \omega_l^r = \frac{-i(l-1)\pm\sqrt{(l-1)(1-lQ_0^4)}}{1+Q_0^2}.
\end{equation} 
\footnote{K. Tanabe has informed us that he is also studying the 
dynamics of charged black holes at large $D$ and has independently 
obtained the result \eqref{scspf}.}

\eqref{scspf}, our final result for the light quasinormal mode frequencies 
associated with shape fluctuations, is correct as stated for $l >1$, 
but requires clarification for in special cases $l=0$ and $l=1$ for the 
reasons we now describe. 

Let us first consider $l=0$. In this case 
\eqref{scspf} predicts the existence of quasinormal modes with 
frequencies  $\omega r_0=0$ and $\omega r_0= \frac{2 i}{1 +Q_0^2}$. 
As noted under \eqref{phconst}, however, modes at $l=0$ are physical only if 
they are time independent. It follows that we have only one mode 
at $l=0$: this mode has $\omega=0$. \footnote{Had the mode with 
$\omega r_0= \frac{2 i}{1 +Q_0^2}$ been physical, it would have represented
an instability, contradicting the well known stability of Reissner Nordstrom 
black holes (atleast of sufficiently small charge) in arbitrary dimensions.}
This zero mode has a simple physical interpretation; it corresponds to an
infinitesimal uniform rescaling of the black hole radius.

Let us now turn to $l=1$. In this case we have a degeneracy of quasinormal 
mode frequencies; both modes have $\omega=0$. The formula \eqref{scspf} was
obtained by assuming harmonic dependence in time and solving for the 
harmonic frequencies, and it is well known that this procedure requires 
modification in the case that the frequencies are degenerate. In order 
to see how this works, we note that the specialization of 
\eqref{decsc} to modes with $l=1$ yields the very simple equation 
\begin{equation}\label{decsclo}
\partial_t^2\delta r  =0.
\end{equation}
It follows that the two solutions to this equation are 
$\delta r= Y_1^m (a_m + b_m t)$ where $a_m$ and $b_m$ are arbitrary 
constants.  These two zero modes also have a simple physical interpretation. 
The mode multiplying $a_m$ is an infinitesimal translation of the black hole, 
while $b_m$ parameterizes an infinitesimal boost of the black hole. Note that
the $m$ labels for $l=1$ scalar spherical harmonics are precisely the labels
for a vector in $(D-1)$ dimensions, as appropriate for translations and boosts. 

As we have mentioned above, \eqref{scspf} is correct as stated for $l \geq 2$. 
It is easily verified \footnote{Note that a Reissner Nordstrom black hole with $Q_0=1$ is extremal at large $D$. 
All regular black holes have $|Q_0|<1$. } that all quasinormal modes with 
$l \geq 2$ have negative imaginary components (and so represent decaying 
fluctuations).

\subsection{The spectrum of velocity fluctuations}\label{spvel}

The fact that the shape fluctuation $\delta r$ obeys the equation of the 
previous subsection ensures that the RHS of \eqref{vecsc} vanishes. The 
velocity fluctuations, $\delta v_a$, are thus effectively constrained to 
obey \eqref{vecsc} with its RHS set to zero. 

The fluctuation field $\delta v$ may be expanded in vector spherical harmonics
\begin{equation}\label{vexp}
\delta v_a = \sum_{l,m} b_{lm} Y_a^{lm} e^{-i \omega^v_l t}
\end{equation}

Let us pause to describe vector spherical harmonics in arbitrary dimension
in more detail. The $l^{th}$ vector spherical harmonic may be obtained as 
a restriction of a vector field on $R^{D-1}$ to the unit sphere. The 
vector field in question is made up as a linear sum of vector valued 
monomials of the form 
$V_{\mu \mu_1 \mu_2 \ldots \mu_l} x^{\mu_1} x^{\mu_2} \ldots x^{\mu_l}$
where $V_{\mu \mu_1 \mu_2 \ldots \mu_l}$ is traceless, symmetric in all of its indices except the first one, and it is zero when it's first index is symmetrized 
with any of the others. In particular, tracing the first index of $b$ with 
any of the others gives zero. 

It follows that each of the vector valued monomials listed above obeys the 
equations 
\begin{equation}\label{fsu}
\nabla. V=0,~~~~~ \nabla^2 V=0
\end{equation}
where the covariant derivatives are taken in the flat space $R^{D-1}$. 
The restriction of each of these vector valued monomials to the unit sphere
yields a vector field tangent to the unit sphere (this is because the 
$r$ component of these vector fields - proportional to the monomial with 
first index dotted with $x^\mu$ - vanishes). Let this vector field be denoted 
by $V$. It is easily verified that $\nabla.V=0$ (where the covariant 
derivative is now taken on the unit sphere). We demonstrate in 
Appendix \ref{vsh} that 
\begin{equation} \label{evsh}
\nabla^2 V=- [(D+l-3)l-1]V
\end{equation}
where, in this equation, $V$ is viewed as a vector field on the unit sphere
and $\nabla$ is the covariant derivative on the unit sphere.

Plugging  the expansion \eqref{vexp} into \eqref{vecsc} and setting the 
coefficient of every independent vector spherical harmonic to zero we obtain,
at leading order in large $D$
\begin{equation}\label{svsp}
\omega^v_l = \frac{-i(l-1)}{1+Q_0^2}.
\end{equation}
This formula agrees with the formula for the spectrum of vector quasinormal 
modes presented in \cite{our} in the limit  
$Q_0\rightarrow 0$.
Reinstating factors of $r_0$ we have 
\begin{equation}\label{svspm}
r_0 \omega^v_l = \frac{-i(l-1)}{1+Q_0^2} ~~~(l=1, 2, 3 \ldots .)
\end{equation}
Note that all the velocity quasinormal modes are pure (negative) 
imaginary, and so represent a ring down that decays without oscillations. 
Vector harmonics with  $l=1$ are zero modes. These modes transform 
in the representation $(1,1,0,0, \ldots 0)$ - i.e. the adjoint 
representation - of $SO(D-1)$ and have a simple physical interpretation. 
These zero modes turn on an infinitesimal spin on the for the black holes, 
i.e. begin to take one along the branch of the large $D$ version of 
Kerr Newman black holes.

\subsection{The spectrum of charge fluctuations}\label{spcharge}

The spectrum of charge fluctuations is governed by the second of 
\eqref{Memeqsc}, which we repeat here for convenience
\begin{equation}
\frac{\nabla^2Q}{\mathcal{K}}-u\cdot\nabla Q = Q\left(\frac{u\cdot\nabla\mathcal{K}}{\mathcal{K}}-u\cdot K\cdot u\right).
\end{equation}
Plugging \eqref{pert} into this equation we obtain the linearized equation
\begin{equation} \label{qlin}
\left(\frac{\nabla^2}{D}-\partial_t\right) \delta Q = Q_0\left(\partial_t^2-\partial_t\left(\frac{\nabla^2}{D}+1\right)\right)\delta r.
\end{equation}

Plugging in the expansion 
\begin{equation}
\delta Q = \sum_{l,m} q_{lm}Y_{lm}(\theta)e^{-i t\omega^Q_l}~~,
\end{equation}
and focusing on the coefficient of $Y_{lm}$ for a particular value of 
$l$, the RHS of \eqref{qlin} is a source term which drives $\delta Q$ at 
the frequency  $\omega_l^r$ given by \eqref{sc1fr}. A source of the form 
\begin{equation}
\delta r=\sum_{l,m}a_{lm}Y_{lm}(\theta)e^{-i\omega^r_l t}
\end{equation}
induces the response 
\begin{equation} \label{psq}
\delta Q_f = \sum_{l,m}a_{lm}\frac{i\omega_l^rQ_0(l-1-i\omega_l^r)}{l-i\omega_l^r}Y_{lm}(\theta)e^{-i\omega^r_l t}~~.
\end{equation}

The most general solution of \eqref{qlin} is given by a linear sum of 
the particular solution \eqref{psq} and the most general solution to the 
homogeneous equation, i.e. to the equation \eqref{qlin} with the RHS set 
to zero. In order to determine the quasinormal frequencies we associated 
with $Q$ oscillations we solve for the frequencies of these homogeneous modes.
This is easily accomplished. Using \eqref{sphericalharm} we find, at 
leading order in large $D$, 
\begin{equation}
-l + i\omega^Q_l = 0,
\end{equation}
which gives the QN frequency for the charge perturbations
\begin{equation}
\omega^Q_l = -il.
\end{equation}
Reinstating factors of $r_0$ we have
\begin{equation}
r_0 \omega^Q_l = -il.
\end{equation}
As in the case of velocity fluctuations, the charge fluctuation quasinormal
modes are pure negative imaginary, and so represent diffusive decay without 
oscillation. $\omega$ vanishes when $l=0$. The corresponding zero mode is 
simply an infinitesimal uniform rescaling of $Q_0$.

\subsection{A consistency check for shape fluctuations}\label{conshape}

The spectrum of shape fluctuations can be rederived starting from 
the equation \eqref{sceq}, i.e. 
\begin{equation}
(1-Q^2)\left[\frac{\nabla^2\mathcal{K}}{\mathcal{K}^2}-\frac{u\cdot\nabla\mathcal{K}}{\mathcal{K}}\right]=(1+Q^2)\left(\frac{u\cdot\nabla\mathcal{K}}{\mathcal{K}}-u\cdot K\cdot u\right).
\end{equation}
The linearized equation is given by
\begin{equation}\label{linsc1}
(1-Q_0^2)\left(\partial_t-\frac{\nabla^2}{D}\right)\left(1+\frac{\nabla^2}{D}\right)\delta r=(1+Q_0^2)\left(\partial_t^2-\partial_t\left(1+\frac{\nabla^2}{D}\right)\right)\delta r.
\end{equation}
Now let's consider the perturbation in membrane shape function to be a particular mode, namely
\begin{equation}
f(t,\theta) = \sum_{l,m} a_{lm} Y_{lm}(\theta)e^{-i\omega_l^r t}~~.
\end{equation}
This turns \eqref{linsc1} into an algebraic equation for a given mode
\begin{equation}
(\omega_l^r)^2 + \frac{2i\omega_l^r(l-1)}{1+Q_0^2} - \frac{1-Q_0^2}{1+Q_0^2}l(l-1) = 0,
\end{equation}
which has roots
\begin{equation}\label{sc1fr}
\omega^r_l = \frac{-i(l-1)\pm\sqrt{(l-1)(1-lQ_0^4)}}{1+Q_0^2}.
\end{equation}
They exactly match with \eqref{scsp}.

Recall that we have argued above that the divergence of \eqref{Memeqsc} 
agrees with \eqref{sceq} for all configurations that preserve an 
SO$(D-p-2)$ isometry. In this subsection we have shown, however, 
the spectrum of shape fluctuations computed from the divergence 
of \eqref{Memeqsc} agrees with the spectrum computed from 
\eqref{sceq} even though arbitrary spherical harmonics do not, in general, 
preserve a large isometry subgroup. The reason this had to work is as follows.
In any spherical harmonic representation there exist special spherical 
harmonics that preserve a large isometry group. It follows from our 
general arguments above the two equations considered in this subsection 
must give the same spectrum of shape fluctuations for these special modes. 
However the equations analyzed in this subsection are geometrical, and 
in particular respect the full SO$(D-1)$ rotational symmetry group of
the background solution, and so generate the same spectrum of oscillations 
for all spherical harmonics in a given representation.

In summary the two equations had to give the same spectrum for {\it some} 
particular elements of the spherical harmonic representation. Rotational 
invariance then forces them to give the same spectrum for all spherical 
harmonics in the same representation, as we actually find.

\section{Discussion} \label{disc}

In this paper we have presented a construction of a large class of solutions
of the Einstein-Maxwell equations. Our solutions are in one to one 
correspondence with the solutions of the equations of a charged, 
nongravitational membrane propagating in flat space according to the 
dynamical equations \eqref{Memeqsc}. 

We have used our membrane equations to generate a prediction 
for the spectrum of light 
quasinormal modes about Reissner-Nordstrom black holes in Einstein-Maxwell 
gravity. As a check on our results it would be useful to independently 
compute these quasinormal mode frequencies, perhaps using the 
gauge invariant formulation of \cite{Kodama:2003kk}.

All of the computations presented in this paper have been performed 
at first nontrivial order in the expansion in $1/D$. It is of great interest
to generalize the computations presented herein to the next order in this 
expansion. Apart from 
determining second order corrections to the membrane equations 
presented in this paper such a computation would allow us to distinguish between 
different geometrical presentations of the first order equations 
(e.g the equation \eqref{sceq} and the divergence of \eqref{Memeqsc}) (see the introduction for a discussion). 

In this paper we have derived equations of membrane dynamics assuming 
that our configuration preserves an $SO(D-p-2)$ isometry. As we have explained 
above, however, our final results are geometrical (in that they make no 
reference to the isometry algebra and treat all dimensions democratically). 
It is possible that the final geometrical equations are valid in more general 
situations, i.e. for configurations that preserve no isometry but perhaps 
obey some other weaker conditions \footnote{We thank A. Strominger for a 
question about this possibility.}. It would be interesting to investigate 
this further.

In order to gain intuition for the membrane equations derived in this paper, 
it would be useful to determine and study the properties of a class of 
simple solutions of these equations. In the upcoming paper \cite{bh} we 
will present a detailed study of stationary solutions to the 
membrane equations \eqref{Memeqsc}. As we have mentioned in the 
introduction, this allows us to make contact between the membrane 
equations presented in this paper and the membrane analysis of 
static and stationary black holes at large $D$ presented in 
\cite{Emparan:2015hwa, Suzuki:2015iha}.

It would also be interesting to follow the lead of \cite{Emparan:2015gva, 
Suzuki:2015axa, Tanabe:2015hda} and attempt to investigate Gregory-Laflamme type instabilities using an appropriate extension of the framework presented in this paper. 

The solutions presented in this paper rapidly approach empty flat space 
away from their event horizons. At every order in the expansion in $1/D$
the gauge field and metric simply vanishes far away from the membrane. 
Non perturbatively in $1/D$ (most likely at order $1/D^D$) we expect 
our membrane motions to excite a Maxwell and gravitational radiation 
field. As this radiation field is the means by which an external observer 
can actually observe the black hole dynamics described in this paper, 
it is of great interest to find the formula that determines this field. 
We hope to return to this question in the near future. 

On a related note, any extended object that consistently sources gravity and 
Maxwell radiation should possess a conserved charge current and stress 
tensor. It would be interesting to find 
all orders formulae (within the $1/D$ expansion) for the charge 
current and stress tensor of the membrane. 

Once all these issues have been settled satisfactorily, it would of course 
be interesting to simulate complicated dynamical processes (e.g. black 
hole collisions) using our membrane equations, and compare our results 
with numerical simulations in $D=4$. Such a comparison would throw light 
on the question of whether the beautiful structures that emerge in black 
hole dynamics at large $D$ are also a useful starting point for a perturbative
expansion for the dynamics of astrophysical black holes.

\vskip .8cm
\section*{Acknowledgments}
We would like to thank A. De and A. Saha for collaboration in the 
initial stages of this work and S. Das, Y. Dandekar, R. Emparan, 
K. Inbasekar, S. Mazumdar, A. Singh,  A. Strominger and S. Trivedi for useful 
discussions. We would also like to thank T. Takayanagi and K. Tanabe 
for comments on a preliminary version of this manuscript. 
S.B. would like to acknowledge the hospitality of the 
University of Barcelona while this work was in progress. S.M. would like 
to acknowledge the hospitality of the Institute of Mathematical Sciences, 
Chennai in the final stages of this work. 
The work of S.B. was supported by an India Israel (ISF/UGC) 
joint research grant. The work of M.M, S.M and S.T was supported 
by a separate India Israel (ISF/UGC) grant, as well as the Infosys Endowment 
for the study of the Quantum Structure of Space Time. We would all also like 
to acknowledge our debt to the people of India for their steady and 
generous support to research in the basic sciences.

\appendix

\section{Reissner-Nordstrom solution in Kerr-Schild coordinates} \label{ks}

The static Reissner-Nordstrom black hole solution is very familiar. 
This solution is most usually presented in Schwarzschild like coordinates.  
In these coordinates the spacetime manifestly Minkowskian  at infinity. 
However the coordinates are singular at the black hole horizon. Let 
${\tilde t}$ be the Schwarzschild time coordinate. The coordinate change
\begin{equation}\label{coc}
d {\tilde t} = dv - \frac{dr}{f(r)},
\end{equation}
recasts the solution as
\begin{equation}\begin{split}\label{bhgfef} 
ds^2&= 2 dv dr - f(r) dv^2 + r^2 d \Omega_{D-2}^2 \,,\\
A&=\sqrt{2} Q \left( \frac{r_0}{r} \right)^{D-3} dv\,,\\
&f(r)=1 - (1+Q^2 c_D) \left( \frac{r_0}{r} \right)^{D-3} + 
c_D Q^2 \left( \frac{r_0}{r} \right)^{2(D-3)},\\
& c_D= \frac{D-3}{D-2}.\\
\end{split}
\end{equation}
In these so called Eddington-Finkelstein coordinates the advantages and 
disadvantages of the Schwarzschild coordinate system are reversed. The black
hole metric is now smooth at the future event horizon. However in the 
limit $r \to \infty$  the spacetime metric $ds^2=2 dv dr -dv^2 + r^2  
d \Omega_{D-2}^2$ is not manifestly Minkowskian. 

The further coordinate change to the `Kerr-Schild' time coordinate $t$ is 
specified by 
\begin{equation} \label{cocn}
dv= dt+dr.
\end{equation}
It is easy to see that the Kerr-Schild time $t$ agrees with the Schwarzschild 
time coordinate at large $r$, but effectively reduces to the Eddington-Finkelstein
time coordinate at the first zero of $f(r)$ (when approached 
from large $r$), i.e. at the outer event horizon. For this reason one 
might anticipate that the black hole solution in Kerr-Schild coordinates 
is both manifestly Minkowskian at large $r$ as well as manifestly smooth 
at the outer future event horizon. A glance at the explicit black hole 
solution, \eqref{bhgfks} is sufficient to convince oneself that this 
is indeed the case.

\section{Relating the geometric form of the metric and gauge field with the answer found in explicit computation}\label{App:compare}
In this appendix we shall present how the different structures and functions appearing in section \ref{efgf} are related to the functions and data appearing in subsection \ref{pertcomp}  (the explicit computation with SO$(d+1)$ invariance).\\
As explained in subsection \ref{suc}, for explicit computation we assumed the following metric for the flat space-time. 
$$ds_{flat}^2=\eta_{MN}dx^M dx^N = \eta_{\mu\nu} dx^\mu dx^\nu + S^2 \Omega_{ij} d\theta^i d\theta^j\,,$$
where $\eta_{\mu\nu} dx^\mu dx^\nu = dw_a dw^a +dS^2$ is the metric in the auxiliary space of $\{x^\mu\}$ (see subsection \ref{auxiliary}) and $\Omega_{ij}$ is the metric of a $d$ dimensional unit sphere ($\{\theta^i\}$ are the angular coordinates along the isometry directions).
 Now because of the isometry the geometric forms will have the following properties.
 \begin{enumerate}
 \item For any geometric vector  ${\mathfrak V}_M$ , the components along the $\{\theta^i\}$ directions will be zero (i.e., ${\mathfrak V}_{\theta^i}=0$).
 \item  Similarly for any geometric tensor ${\mathfrak H}_{MN}$ 
 $${\mathfrak H}_{\mu\theta^i} =0  ~~\text{and}~~ {\mathfrak H}_{\theta^i\theta^j}\propto\Omega_{ij}\,.$$
 \item As explained before,  apart from $n_\mu$ and $u_\mu$ there is one more special vector in the auxiliary space:  $Z_M dx^M = Z_\mu dx^\mu = \frac{dS}{S} - \left(\frac{n_S}{S}\right) n_\mu dx^\mu$ . Using these $Z$ one-form we can further decompose any geometric vector and tensor scalar, vector and tensor of the SO$(p)$ isometry group in the $(p+3)$ dimensional auxiliary space.

\end{enumerate}
Using these properties we can  translate the results in the geometric form to the language of `auxiliary space'. For most of the functions, the translation is  straightforward and we present the dictionary in Table \ref{table:geotonongeo1} and Table \ref{table:geotonongeo2}. 
In Table \ref{table:geotonongeo1} we present how the three scalar, two vector and one tensor function appearing in the geometric form of the metric and gauge field (\eqref{metgform}) decompose into seven scalar, three vector and one tensor function appearing in equation (\eqref{ukc}).
  Then in table \ref{table:geotonongeo2} we decompose the geometric data in terms of the non-geometric ones (with SO$(d+1)$ invariance) used for explicit computation.

However for for $\delta\phi$, i.e. the fluctuation in the radius of the $d$ dimensional sphere, the translation rule becomes a bit more subtle to be presented in a table. For convenience we shall explain it separately in subsection \ref{dphi}.\\

\begin{table}[ht]
\vspace{0.5cm}
\centering 
\begin{tabular}{|c| c| } 
\hline 
 Explicit Computation&Geometric Form\\
\hline
\hline
$\left(\frac{1}{D}\right)S_{(VV)}$& $H^{(S)}  + {\cal O}\left(\frac{1}{D}\right)^2$\\
\hline
$\left(\frac{1}{D}\right)S_{(AV)}$&$A^{(S)}+ {\cal O}\left(\frac{1}{D}\right)^2$\\
\hline
$\left(\frac{1}{D}\right)S_{(Vz)}$& $Z^\mu H^{(V)}_\mu+ {\cal O}\left(\frac{1}{D}\right)^2$\\
\hline
$\left(\frac{1}{D}\right)S_{(zz)}$& $Z^\mu Z^\nu H^{(T)}_{\mu\nu}+ {\cal O}\left(\frac{1}{D}\right)^2$\\
\hline
$\left(\frac{1}{D}\right)S_{(Az)}$ &$Z^\mu A^{(V)}_\mu+ {\cal O}\left(\frac{1}{D}\right)^2$\\
\hline
$\left(\frac{1}{D}\right)S_{(Tr)}$&$\left(\frac{1}{p}\right)P^{\mu\nu} H^{(T)}_{\mu\nu} + H^{(Tr)}+ {\cal O}\left(\frac{1}{D}\right)^2$\\
\hline
\hline
$\left(\frac{1}{D}\right)V^{(V)}_\mu$&$P_{\mu}^{\nu}H^{(V)}_\nu+ {\cal O}\left(\frac{1}{D}\right)^2$\\
\hline
$\left(\frac{1}{D}\right)V^{(z)}_\mu$&$P_{\mu}^{\alpha}Z^\beta H^{(T)}_{\alpha\beta}+ {\cal O}\left(\frac{1}{D}\right)^2$\\
\hline
$\left(\frac{1}{D}\right)V^{(A)}_\mu$&$P_{\mu}^{\nu}A^{(V)}_\nu+ {\cal O}\left(\frac{1}{D}\right)^2$\\
\hline
\hline
$\left(\frac{1}{D}\right)T^{\mu\nu}$&$ P^{\mu\alpha} P^{\nu\beta}\left[H^{(T)}_{\alpha\beta} -\left(\frac{\mathfrak g_{\alpha\beta}}{p} \right)P^{\nu_1\nu_2} H^{(T)}_{\nu_1\nu_2} \right]+ {\cal O}\left(\frac{1}{D}\right)^2$\\
\hline
\end{tabular}\vspace{.5cm}
\caption{Here we relate how the functions appearing in equation \eqref{ukc} are related to the geometric form of the metric and the gauge field as in equation \eqref{metgform} }
\label{table:geotonongeo1} 
\end{table}
\noindent

\begin{table}[ht]
\vspace{0.5cm}
\centering 
\begin{tabular}{|c| c| } 
\hline 
Geometric data&Data used in computation\\
\hline
\hline
${\mathfrak S}_{(1)} $&$\left(\frac{S}{n_S}\right)\left[\frac{{\mathfrak s}^{(5)}}{Q} -{\mathfrak s}^{(1)} +\left(\frac{S}{n_S}\right){\mathfrak s}^{(2)}\right]$\\
\hline
\hline
${\mathfrak S}_{(2)}$&$\left(\frac{S}{n_S}\right)\left[{\mathfrak s}^{(1)} -\left(\frac{S}{n_S}\right){\mathfrak s}^{(2)}\right]$\\
\hline
\hline
$Z_\mu {\mathfrak V}_{(1)}^\mu$&$\left(\frac{S}{n_S}\right)^2{\mathfrak s}^{(6)} +{\mathfrak s}^{(1)} -\left(\frac{1-n_S^2}{S\times  n_S}\right)^2$\\
\hline
\hline
$Z_\mu {\mathfrak V}_{(2)}^\mu$&$\left(\frac{S}{n_S}\right)^2{\mathfrak s}^{(6)} -{\mathfrak s}^{(1)} -\left(\frac{1-n_S^2}{S\times  n_S}\right)^2$\\
\hline
\hline
$Z_\mu Z_\nu {\mathfrak T}^{\mu\nu}$&${\mathfrak s}^{(6)} -\left(\frac{n_S}{S}\right){\mathfrak s}^{(2)}$\\
\hline
\hline
$P_{\mu\nu} {\mathfrak T}^{\mu\nu}$&${\mathfrak s}^{(3)} -{\mathfrak s}^{(4)}$\\
\hline
\hline
$P_{\mu\nu} {\mathfrak V}_{(1)}^\nu$&$ \left(\frac{S}{n_S}\right)^2\left[{\mathfrak v}_\mu^{(5)}+\left(\frac{n_S}{S}\right){\mathfrak v}_\mu^{(2)}\right] $ \\
\hline
\hline
$P_{\mu\nu} {\mathfrak V}_{(2)}^\nu$&$ \left(\frac{S}{n_S}\right)^2\left[{\mathfrak v}_\mu^{(5)}-\left(\frac{n_S}{S}\right){\mathfrak v}_\mu^{(2)}\right] $ \\
\hline
\hline
$Z_{\nu} P_{\alpha\mu}{\mathfrak T}^{\alpha\nu} $&${\mathfrak v}_\mu ^{(5)} -{\mathfrak v}_\mu ^{(3)}- \left(\frac{n_S}{S}\right){\mathfrak v}_\mu ^{(1)}$\\
\hline
\end{tabular}\vspace{.5cm}
\caption{Decomposition of geometric data in the special case of SO$(d+1)$ symmetry  in terms of the data in auxiliary space used for explicit computation }
\label{table:geotonongeo2} 
\end{table}
\noindent

Using the tables \ref{table:geotonongeo1} and \ref{table:geotonongeo2} and the argument presented in subsection \ref{dphi}  we could easily see that
if we specialize the metric and the gauge field as presented in equations \eqref{metgform}, \eqref{gaugegeo}, \eqref{metricgeo1} and \eqref{metricgeo2} to  SO$(d+1)$  isometry (where $d= D-p-3)$ ,  they indeed reduce to the explicit solution presented in subsection \ref{pertcomp} upto correction of ${\cal O}\left(\frac{1}{D}\right)^2$.\\

\subsection{Relating $\delta\phi$ to the geometric forms}\label{dphi}
  Note that any geometric tensor will have some nonzero components along the isometry directions and also because of symmetry the components must be proportional to the metric of the $d$ dimensional unit sphere. We can explicitly compute this proportionality factor which will be directly related to $\delta\phi$. \\
Consider the  traceless tensor  $H^{(T)}_{MN}$ appearing in equation \eqref{metgform} and suppose $H^{(T)}_{\theta^i\theta^j}=S^2 {\mathfrak H}~\Omega_{ij}$ where ${\mathfrak H}$ is some scalar function. Then it follows that
 \begin{equation}\label{traceform1}
 \begin{split}
 &0= \eta^{MN}H^{(T)}_{MN} = \eta^{\mu\nu}H^{(T)}_{\mu\nu} +\frac{\Omega^{ij}}{S^2} H^{(T)}_{\theta^i\theta^j}=\eta^{\mu\nu}H^{(T)}_{\mu\nu}+d\times{\mathfrak H}\\
 \Rightarrow ~&{\mathfrak H} =- \frac{\eta^{\mu\nu}H^{(T)}_{\mu\nu}}{D-p-3} .
 \end{split}
 \end{equation}
 Similarly consider the tensor $H^{(Tr)}{\cal P}_{MN}$ appearing in \eqref{metgform}). Since we know that $n_{\theta^i} =u_{\theta^i}=0$, the nonzero components of this tensor along the isometry directions are simply given by
 \begin{equation}\label{traceform2}
 \begin{split}
H^{(Tr)}{\cal P}_{\theta^i\theta^j} = H^{(Tr)}S^2\Omega_{ij}\,.
 \end{split}
 \end{equation}

From equation \eqref{traceform1} and \eqref{traceform2} and the definition of $\delta\phi$ (recall that the fluctuation in the radius of the $d$ dimensional sphere $=S^2\delta\phi$),  it follows that 
\begin{equation}\label{trform}
\delta\phi= H^{(Tr)} + {\mathfrak H}=H^{(Tr)} - \left(\frac{1}{D-p-3}\right)\eta^{\mu\nu} H^{(T)}_{\mu\nu}\,.
\end{equation}
The second term in the RHS of equation \eqref{trform} is of ${\cal O}\left(\frac{1}{D}\right)^2$ since by construction $H^{(T)}_{\mu\nu}$ starts at ${\cal O}\left(\frac{1}{D}\right)$ .
Now from explicit computation we know that $\delta\phi$ is of ${\cal O}\left(\frac{1}{D}\right)^2$ (see equation \eqref{expansionPhi}) . Then it immediately follows that $H^{(Tr)}$ also must start from terms of ${\cal O}\left(\frac{1}{D}\right)^2$.

We could explicitly compute the second term in RHS of \eqref{trform} in terms of the functions appearing in equation \eqref{ukc}. Note that $\eta^{\mu\nu}$ could be expanded  as
$$\eta^{\mu\nu} = n^\mu n^\nu - u^\mu u^\nu + \frac{S^2}{1-n_S^2} Z^\mu Z^\nu + P^{\mu\nu} + {\cal O}\left(\frac{1}{D}\right).$$
Using this expansion of $\eta^{\mu\nu}$  and the translation rules as given in table \ref{table:geotonongeo1} we find
\begin{equation}\label{hhten}
\left(\frac{1}{D-p-3}\right)\eta^{\mu\nu} H^{(T)}_{\mu\nu}=  \frac{1}{D^2}\left[\left(\frac{1-n_S^2}{S^2}\right)S_{zz} + p \times S_{(Tr)}\right]+{\cal O}\left(\frac{1}{D}\right)^3.
\end{equation}
In equation \eqref{hhten} we have used the fact that $H^{(Tr)}$ is of  ${\cal O}\left(\frac{1}{D}\right)^2$ and by construction $u^\mu  H^{(T)}_{\mu\nu} = n^\mu  H^{(T)}_{\mu\nu} =0$.
Substituting equation \eqref{hhten} in equation \eqref{trform} we found that 
$$H^{(Tr)} = \frac{1}{D^2}\left[\delta\phi^{(2)} + \left(\frac{1-n_S^2}{S^2}\right)S_{zz} + p \times S_{(Tr)}\right]+{\cal O}\left(\frac{1}{D}\right)^3.$$ 

Now from equation \eqref{expansionPhi} it directly follows that 
$$H^{(Tr)} = {\cal O}\left(\frac{1}{D}\right)^3.$$

\section{Relating equations of motion expressed in different forms} \label{app:eqnrel}
In this section we shall first  state a set of algebraic identities that are true upto corrections of ${\cal O}\left(\frac{1}{D}\right)$. Using these identities we could easily show that the equations of motion as derived in subsection \ref{eomregh} (\eqref{relationss} and \eqref{relationsv}) are equivalent to equations \eqref{chdeqn1} and \eqref{chdeqn2}.
\begin{enumerate}
\item Identity-1:
\begin{equation}\label{identity1}
\begin{split}
\tilde{V}_\perp\cdot K\cdot u &= u\cdot K\cdot \tilde{V}_\perp
=[(u\cdot\partial) n]\cdot \tilde{V}_\perp\\
&=-[(u\cdot\partial) \tilde{V}_\perp]\cdot n~~~~~~~~~~\text{Since $ n\cdot \tilde{V}_\perp =0$}\\
&=\left(\frac{n_S}{S} \right)[(u\cdot\partial) u]\cdot n+(u\cdot\partial) \left(\frac{n_S}{S} \right)+{\cal O}\left(\frac{1}{D}\right) ~~~~\text{Since $ u\cdot dS ={\cal O}\left(\frac{1}{D}\right)$}\\
&=-\left(\frac{n_S}{S} \right)(u\cdot K\cdot u)+(u\cdot\partial) \left(\frac{n_S}{S} \right)+{\cal O}\left(\frac{1}{D}\right) ~~~~\text{Since $ u\cdot n =0$}.\\
\end{split}
\end{equation}
Here $\tilde{V}_\perp = Z -\left(\frac{n_S}{S}\right) u = \frac{dS}{S} -\left(\frac{n_S}{S}\right)(n+u)=\left(\frac{S}{n_S}\right)(X-u).$
\item Identity-2:
\begin{equation}\label{identity2}
\begin{split}
\tilde{V}_\perp\cdot\partial u\cdot Z
=&-\tilde{V}_\perp\cdot\partial Z\cdot u +{\cal O}\left(\frac{1}{D}\right)~~~~\text{since $u\cdot Z ={\cal O}\left(\frac{1}{D}\right)$}\\
=&\left(\frac{n_S}{S}\right)\tilde{V}_\perp\cdot K\cdot u+{\cal O}\left(\frac{1}{D}\right)~~~~\text{since $u\cdot dS = {\cal O}\left(\frac{1}{D}\right),~~u\cdot n=0$}.\\
\end{split}
\end{equation}
\item Identity-3:
\begin{equation}\label{identity3}
\begin{split}
u\cdot\partial u\cdot Z
=&-u\cdot\partial Z\cdot u +{\cal O}\left(\frac{1}{D}\right)~~~~\text{since $u\cdot Z ={\cal O}\left(\frac{1}{D}\right)$}\\
=&\left(\frac{n_S}{S}\right)u\cdot K\cdot u+{\cal O}\left(\frac{1}{D}\right)~~~~\text{since $u\cdot dS = {\cal O}\left(\frac{1}{D}\right),~~u\cdot n=0$}.\\
\end{split}
\end{equation}
\item Identity-4:
\begin{equation}\label{identity4}
\begin{split}
&\left(\frac{S}{n_S}\right)\tilde{V}_\perp\cdot\partial (u-n) \cdot Z\\
=~&\tilde{V}_\perp\cdot K\cdot u -\left(\frac{S}{n_S}\right)\tilde{V}_\perp\cdot K\cdot Z +{\cal O}\left(\frac{1}{D}\right)~~~~~~\text{Using \eqref{identity2}}\\
=~&-\left(\frac{S}{n_S}\right)\tilde{V}_\perp\cdot K\cdot \tilde{V}_\perp+{\cal O}\left(\frac{1}{D}\right)~~~~~~\text{Since $Z= \tilde{V}_\perp + \frac{n_S}{S} u$}.
\end{split}
\end{equation}
\item Identity-5:
\begin{equation}\label{identity5}
\begin{split}
&\left[u\cdot\partial u-\left(\frac{S}{n_S}\right)Z\cdot\partial n \right]\cdot Z\\
=~&\left(\frac{n_S}{S}\right)u\cdot K\cdot u -\left(\frac{S}{n_S}\right)Z\cdot K\cdot Z+{\cal O}\left(\frac{1}{D}\right)~~~~~~\text{Using \eqref{identity3}}\\
=~&-\left(\frac{S}{n_S}\right)\tilde{V}_\perp\cdot K\cdot \tilde{V}_\perp-2(Z\cdot K\cdot u) +2\left(\frac{n_S}{S}\right)u\cdot K\cdot u+{\cal O}\left(\frac{1}{D}\right)\\
=~&-\left(\frac{S}{n_S}\right)\tilde{V}_\perp\cdot K\cdot \tilde{V}_\perp-2(\tilde{V}_\perp\cdot K\cdot u) +{\cal O}\left(\frac{1}{D}\right).\\
\end{split}
\end{equation}

Using \eqref{identity4}and \eqref{identity5} we could very easily compute  the projection of \eqref{chdeqn1} along $Z$ direction. It turns out to be the following,
\begin{equation}\label{zcomp}
\begin{split}
0=&\bigg[(u-X)\cdot\partial O  - Q^2 (u\cdot\partial)u-Q^2(X\cdot K) \bigg]\cdot Z+\left(\frac{n_S}{S}\right)(1-Q^2)(X\cdot Z)\\
=& -(1-Q^2)\left[\left(\frac{S}{n_S}\right)\tilde{V}_\perp\cdot K\cdot \tilde{V}_\perp - Z\cdot Z\right] +2Q^2(\tilde{V}_\perp\cdot K\cdot u) +{\cal O}\left(\frac{1}{D}\right)\\
=& -\left(\frac{S}{n_S}\right)(1-Q^2)\left[\tilde{V}_\perp\cdot K\cdot \tilde{V}_\perp - \left(\frac{n_S(1-n_S^2)}{S^3}\right)\right] +2Q^2(\tilde{V}_\perp\cdot K\cdot u) +{\cal O}\left(\frac{1}{D}\right).\\
\end{split}
\end{equation}
Equation \eqref{zcomp} is simply  equal to $\left[-\frac{S}{n_S} (1-Q^2)\right]$ times the first equation in \eqref{relationss}.
Second equation of \eqref{relationss} follows once we substitute \eqref{identity1} in equation \eqref{chdeqn2}. 

\end{enumerate}

\section{Notation and translation} \label{translation}

Through this paper we have had occasion to work with functions (like $\rho$ 
and $Q$)  and one-form or vector fields (like $u$ and 
$n= \frac{d \rho}{|d \rho|}$) that live in flat $D$ dimensional space. 
We also often deal with functions and one-forms that live on the 
the membrane world volume. Through the paper we use the dummy indices 
$M, N \ldots$ to denote coordinates in the embedding flat $D$ dimensional spacetime, and 
the indices $A, B \ldots$ to denote coordinates on the membrane world volume. 
$M, N$ indices run over $D$ values, while $A, B$ indices run over $D-1$ 
values.

In the computational part of this paper we have assumed that our spacetimes 
and membrane world volumes both preserve an SO$(D-p-2)$ isometry group. 
It follows that the spacetime metric takes the form 
\begin{equation}\label{stm}
ds^2=g_{\mu\nu}dx^\mu dx^\nu+ e^\phi d \Omega_{d}^2,
\end{equation}
where $\mu, \nu$ run over $p+3$ values and $g_{\mu\nu}$ and $\phi$ are 
functions only of $x^\mu$. We will often use the notation 
$$e^\phi= S^2.$$
In a similar manner the metric on the membrane 
world volume takes the form 
\begin{equation}\label{memm}
ds^2=g_{a b}dx^a dx^b+ e^\phi d \Omega_{d}^2,
\end{equation}
where $a, b$ run over $p+2$ values. As all spacetime vector and scalar 
fields also preserve SO$(d+1)$, for computational purposes it is sometimes 
useful to view these fields as living on the reduced $p+3$ dimensional 
manifold 
$$ds^2=g_{\mu\nu}dx^\mu dx^\nu,$$ 
(in the case of bulk fields) 
and 
$$ds^2=g_{ab} dx^a dx^b,$$
(in the case of fields that live on the membrane world volume). We use the 
symbols $\nabla_M$ and $\nabla_A$ to denote covariant derivatives on all 
of spacetime (or all of the membrane world volume) and ${\tilde \nabla}_\mu$
and ${\tilde \nabla}_a$ to denote fields on reduced spacetime (or the reduced
membrane world volume). 

Consider a vector field $v^M$ defined on all of flat space. If we assume that
$v^M$ preserves SO$(d)$ invariance, it is easily verified that 
\begin{equation}\label{dvvec}
\nabla_M v^M= d \frac{v.{\tilde \nabla }S}{S} + {\tilde \nabla}_\mu v^\mu.
\end{equation}
In a similar manner, if $v^A$ is a vector field on the membrane then 
\begin{equation}\label{dvvec2}
\nabla_A v^A= d \frac{v.{\tilde \nabla} S}{S} + {\tilde \nabla}_a v^a.
\end{equation}

If $\psi$ is a scalar field in spacetime or on the membrane then it is easily 
verified that 
\begin{equation}\label{dvvec3}
\nabla^2 \psi= d \frac{{\tilde \nabla} S. {\tilde \nabla} \psi}{S} + 
{\tilde \nabla}^2 \psi,
\end{equation}
(where $dS$ is regarded as a one-form in either spacetime or on the membrane 
depending on the space on which $\nabla^2$ is evaluated). Note that $dS$ on 
the membrane world volume is simply $dS$ in spacetime, projected onto the 
membrane world volume.

Finally if $v$ is a vector field in either spacetime or the membrane world 
volume then 
\begin{equation}\label{dvvec4}
\nabla^2 v = d \left( \frac{{\tilde \nabla } S.{\tilde \nabla} v}{S} -  
dS ~v.{\tilde \nabla} S \right)  + 
 {\tilde \nabla}^2 v.
\end{equation}

To end this section we note that the extrinsic curvature tensor for the 
membrane world volume takes the form 
$$K_{A B}= (K_{\mu\nu} , \frac{n^S}{S} \Omega_{ij}),$$
where $\theta^i$ are angles on the unit $d$ sphere and 
$\Omega_{ij}$ is a metric on this space.

\section{Eigenvalues of the Laplacian for Vector Spherical Harmonics} 
\label{vsh}

In this Appendix we evaluate the eigenvalue of the Laplacian acting on 
the $l^{th}$ vector spherical harmonic. This spherical harmonic was 
defined in terms of the restriction of a collection of vector valued 
monomials to the unit sphere in  subsection \ref{spvel}.

Our strategy is to evaluate the 
Laplacian of $V$ - viewed as a vector valued monomial in $R^{D-1}$ - 
in spherical polar coordinates, and use the fact that this Laplacian vanishes
(see subsection \ref{spvel}) to evaluate the Laplacian of the same vector 
field restricted to the unit sphere. 

Consider any divergenceless vector field on $R^{D-1}$ with vanishing radial 
component, i.e. $V_r=0$. Using explicit expressions for the 
Christoffel symbols for flat space in polar coordinates we find 
\begin{equation}
\begin{split}
\nabla_rV_r &= 0,\\
\nabla_rV_a &= \partial_rV_a-\frac{V_a}{r},\\
\nabla_aV_r &= \frac{V_a}{r},\\
\nabla_aV_b &= \hat{\nabla}_aV_b,
\end{split}
\end{equation}
where $\hat{\nabla}$ denotes the covariant derivative taken on a unit sphere.
\\
We will now use these results to evaluate $\nabla^2 V$ on $R^{D-1}$ in 
spherical polar coordinates. The result of this computation depends on the 
free index in this equation. Let us first consider the case with the 
free index equal to $r$. In this case 
\begin{equation}
\begin{split}
\nabla^2V_r &= \nabla_r(\nabla_rV_r)+\frac{1}{r^2}g^{ab}\nabla_a\nabla_bV_r,\\
&= \frac{1}{r^2}\hat{\nabla}_a\hat{\nabla}^aV_r - \frac{1}{r^2}\hat{\nabla_a}{V^a},\\
&=0.
\end{split}
\end{equation}
In other words the vanishing of the $r$ component of $\nabla^2 V$ is just a 
triviality - it follows as an identity upon assuming $V_r=0$ and 
$\nabla.V=0$. 

Let us now turn to the more interesting case of the free index being an angular direction on the unit sphere. In this case\\
\begin{equation}
\begin{split}
\nabla^2V_c =& \nabla_r(\nabla_rV_c)+\frac{1}{r^2}g^{ab}\nabla_a\nabla_bV_c,\\
=& \partial_r\left(\partial_rV_c-\frac{V_c}{r}\right)-\Gamma^a_{rc}\left(\partial_rV_a-\frac{V_a}{r}\right)+\frac{1}{r^2}\hat{\nabla}_a\hat{\nabla}^aV_c\\ &+ \Gamma^a_{ar}\left(\partial_rV_c-\frac{V_c}{r}\right)+\frac{1}{r^2}\Gamma^r_{ac}\frac{A^a}{r},\\
=& \partial_r\left(\partial_rV_c-\frac{V_c}{r}\right)-\frac{1}{r}\left(\partial_rV_c-\frac{V_c}{r}\right)+\frac{1}{r^2}\hat{\nabla}_a\hat{\nabla}^aV_c\\ &+ \frac{D-2}{r}\left(\partial_rV_c-\frac{V_c}{r}\right)-\frac{V_c}{r^2}~~.\\
\end{split}
\end{equation}
Let us now specialize to $V_c$ is the vector field corresponding to the 
$l^{th}$ vector spherical harmonic. In this case $V_c \propto r^{l+1}$. 
Using this fact and $\nabla^2 V_c = 0$ we get
\begin{equation}
-\frac{1}{r^2}\hat{\nabla}^2V_c = (l(l+1)-l-l+(D-2)l-1)V_c = [(D+l-3)l-1]V_c.
\end{equation}

\bibliography{mempap}{}
\bibliographystyle{JHEP}

\end{document}